\documentclass[openacc]{rstransa}
\usepackage{float}\usepackage{graphicx}
\usepackage{subfig}
\usepackage{dcolumn}
\usepackage{bm}
\usepackage{enumitem}
\usepackage{bbold}
\usepackage{dsfont}
\usepackage{comment}
\usepackage[dvipsnames]{xcolor}
\usepackage{amsmath,amsfonts,amsthm,bm} 
\usepackage[normalem]{ulem}

\newcommand\ethr{e_\text{thresh}}
\newcommand\domain{\Omega}  
\newcommand\setAone{\mathcal{A}}
\newcommand\setBtwo{\mathcal{B}_2}
\newcommand\setBo{\mathcal{B}_0}
\newcommand\surfS{\mathcal{S}}
\newcommand\setA{\mathcal{A}}
\newcommand\setB{\mathcal{B}}
\newcommand\hA{h_{\setA}}
\newcommand\hAone{h_{\setAone}}
\newcommand\hB{h_{\setB}}
\newcommand\hBo{h_{\setBo}}
\newcommand\hBtwo{h_{\setBtwo}}
\newcommand\hS{h_{\surfS}}
\newcommand\hleft{h_\text{left}}
\newcommand\hright{h_\text{right}}
\newcommand\hone{h_0}
\newcommand\taud{\tau_d}
\newcommand\taus{\tau_s}
\newcommand\Pid{\Pi_d}
\newcommand\Pis{\Pi_s}
\newcommand\oT{\overline{T}}
\newcommand\pAMS{\hat{p}}
\newcommand\pMC{\hat{p}_\text{MC}}
\newcommand\survival[1]{S(#1)}
\newcommand\survivalphi[1]{S_\phi(#1)}

\begin{document}
\title{Extreme events in transitional turbulence}%

\author{
S\'ebastien Gom\'e$^{1}$, Laurette S. Tuckerman$^{1}$ and Dwight Barkley$^{2}$}

\address{$^{1}$ Laboratoire de Physique et M\'ecanique des Milieux H\'et\'erog\`enes, CNRS, ESPCI Paris, PSL Research
University, Sorbonne Universit\'e, Universit\'e de Paris, Paris 75005, France\\
$^{2}$  Mathematics Institute, University of Warwick, Coventry CV4 7AL, United Kingdom}

\subject{fluid mechanics, mathematical modelling, statistical physics}


\corres{Dwight Barkley\\
\email{D.Barkley@warwick.ac.uk}}

\begin{abstract}

Transitional localised turbulence in shear flows is known to either decay to an absorbing laminar state or to proliferate via splitting. The average passage times from one state to the other depend super-exponentially on the Reynolds number and lead to a crossing Reynolds number above which proliferation is more likely than decay. In this paper, we apply a rare event algorithm, Adaptative Multilevel Splitting (AMS), to the deterministic Navier-Stokes equations to study transition paths and estimate large passage times in channel flow more efficiently than direct simulations. We establish a connection with extreme value distributions and show that transition between states is mediated by a regime that is self-similar with the Reynolds number. The super-exponential variation of the passage times is linked to the Reynolds-number dependence of the parameters of the extreme value distribution. Finally, motivated by instantons from Large Deviation Theory, we show that decay or splitting events approach a most-probable pathway.

\end{abstract}

\maketitle


\section{Introduction}

The route to turbulence in many wall-bounded shear flows is a spatiotemporal process that results from the interplay between the tendency for turbulence to decay or for it to proliferate. Individual decay and proliferation events occur extremely rarely near the critical Reynolds number for the onset of sustained turbulence, and this makes measuring, let alone understanding the onset of turbulence in these flows both fascinating and challenging. In this paper we investigate these rare events. 

\begin{figure}[h!]
    \centering
 \includegraphics[width=\columnwidth]{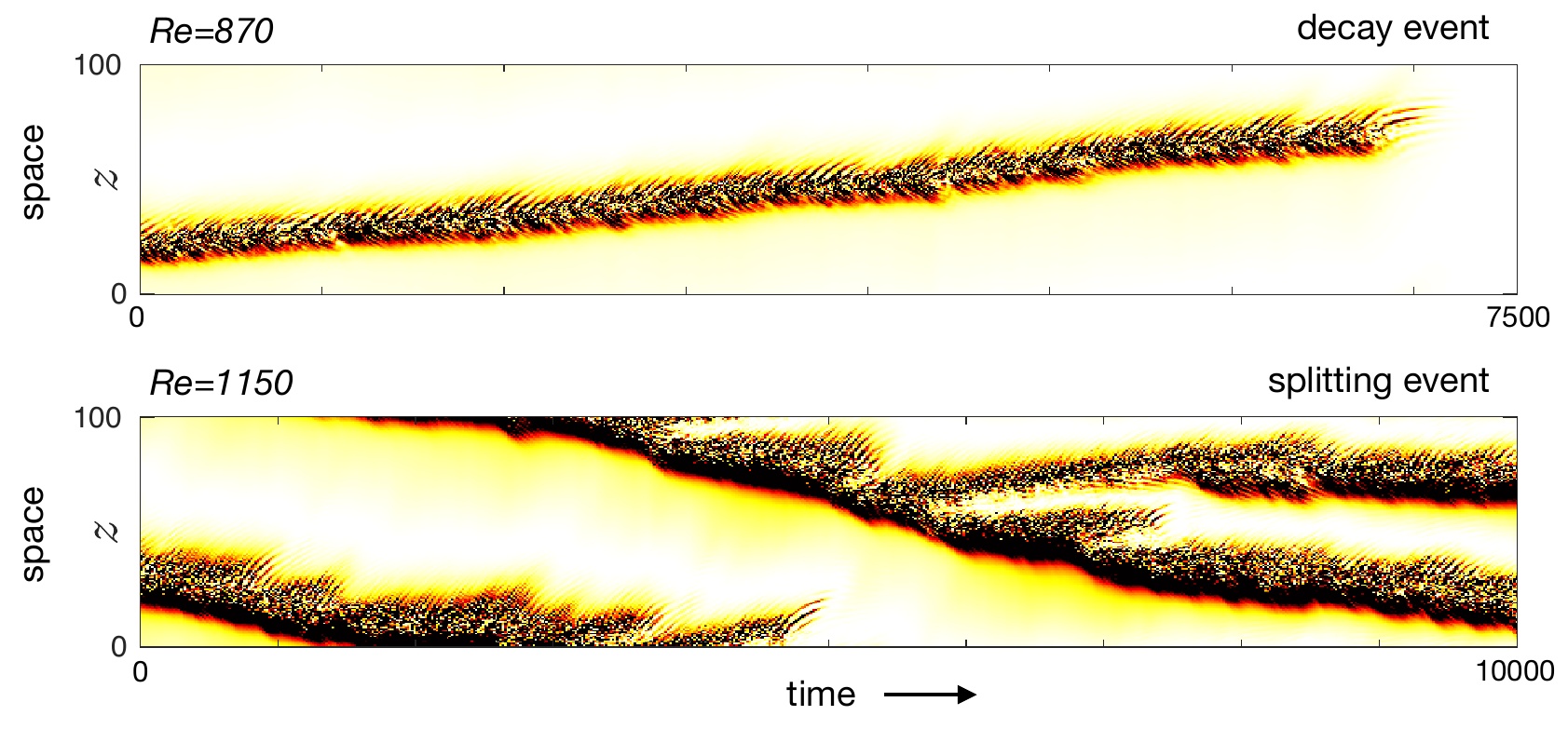} 
    \caption{Evolution of turbulence in channel flow at two different Reynolds numbers. Turbulence is seen as black and is localised to only a portion of space. White corresponds to laminar (or nearly laminar) flow. The motion of the turbulent patch is seen in a frame of reference moving with the mean flow in the channel and the system is periodic in spatial coordinate $z$. At $Re=870$ the localised band of turbulence maintains an approximately constant width and intensity for a considerable time and then abruptly transitions to laminar flow in a decay event.  At $Re=1150$ the localised turbulent band is wider and noticeably asymmetric. In this case the band splits into two bands. In the vicinity of $Re = 1000$, both of these key events become extremely rare and the mean exit time from the one-band state becomes very large.
    Results are obtained by a numerical simulation in an oblique domain represented in Figure ~\ref{fig:domain}.
    }
    \label{fig:intro_time}
\end{figure}

\begin{figure}
    \centering
 \includegraphics[width=\columnwidth]{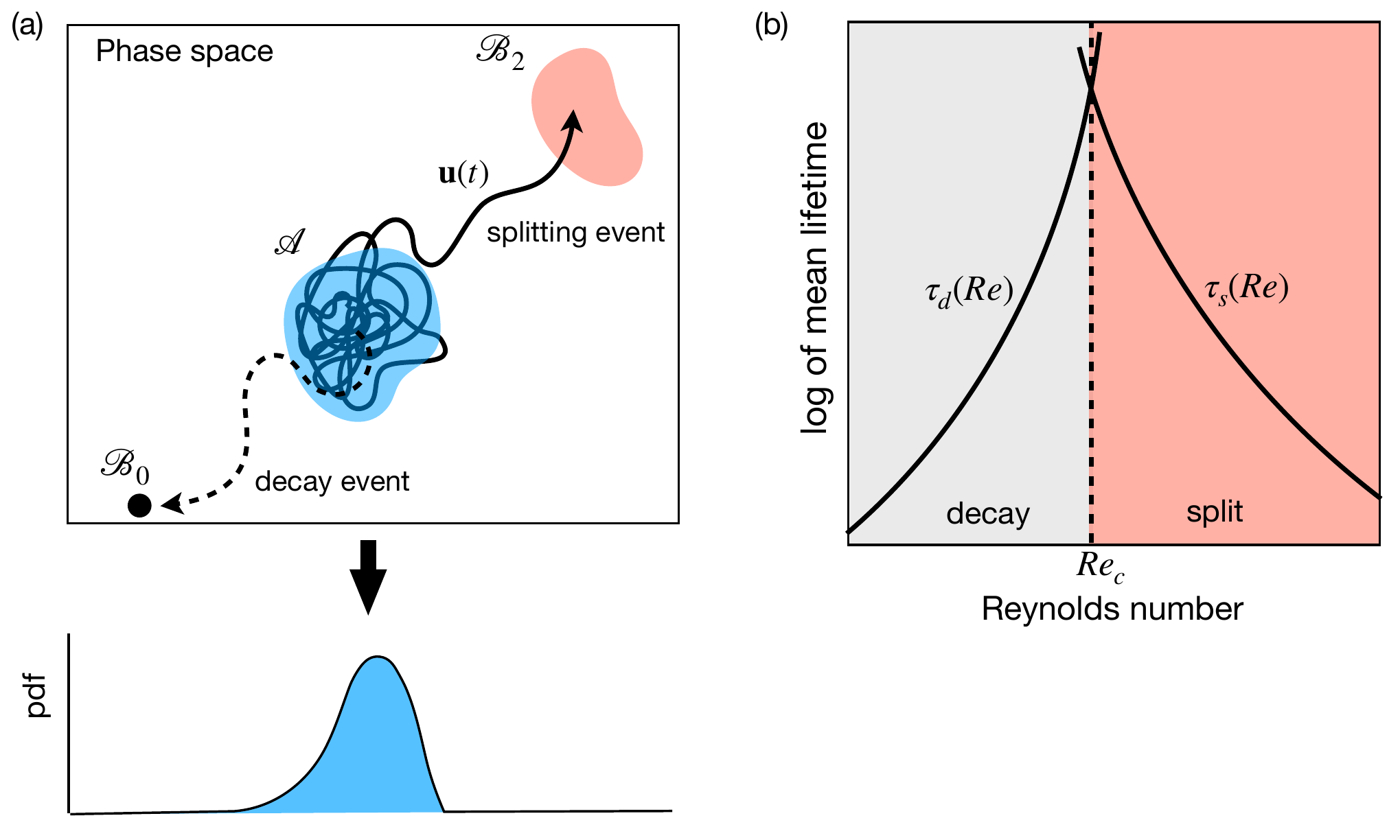} 
    \caption{(a) Illustration of the phase space of the Navier-Stokes equations. Time evolving flow fields $\mathbf{u}(t)$ are seen as trajectories. The one-band state corresponds to a region $\setA$ in the phase space in which trajectories $\mathbf{u}(t)$ spend considerable time before exiting and transitioning either to laminar flow $\setBo$ or to the two-band state $\setBtwo$. 
    The fluctuations of observables, such as the turbulence fraction, are described by extreme value distributions.  
    (b) Schematic showing the dependence of mean lifetimes on Reynolds number, $Re$. Lifetimes vary super-exponentially with $Re$, with $\taud$ increasing and $\taus$ decreasing with $Re$. The timescales cross at a critical value $Re_c$. Below $Re_c$, decay occurs more frequently while above $Re_c$, splitting occurs more frequently.
  }
\label{fig:intro_pp}
\end{figure}

Figure~\ref{fig:intro_time} illustrates individual decay and proliferation (splitting) events of interest. These have been obtained from numerical simulations of pressure-driven flow in a channel. The spatio-temporal diagrams of figure~\ref{fig:intro_time} display the evolution of such localised turbulent bands at two Reynolds numbers. Simulations begin after some initial equilibration time. It can be seen that the one-band state is metastable -- it persists for significant time before transitioning to another state, either laminar flow, as in the upper panel, or a two-band state, as in the lower one. 
The corresponding phase-space picture for the governing Navier-Stokes equations is sketched in figure~\ref{fig:intro_pp}. Trajectories spend a significant time in a region of phase space associated with a single turbulent band, $\setA$, before exiting the region and going to laminar flow or to the two-band state. 
Repeated simulations starting from one-band states (in the region $\setA$) show that the exit times are distributed exponentially, so that decay and splitting events are effectively governed by a memoryless, Poisson process. See  \cite{faisst2004sensitive,eckhardt2007turbulence,avila2010transient,avila2011onset,shi,gome2020statistical} and references therein.

A typical study consists of the following. For each value of the Reynolds number, $Re$, a large number of events is generated, from which the mean lifetime is determined by averaging the lifetimes observed in the sample events. This is the Monte Carlo approach. The process is repeated for a range of $Re$ to obtain the mean lifetimes to decay $\taud(Re)$ and to split $\taus(Re)$.
These lifetimes are observed to depend super-exponentially on Reynolds number as sketched in figure~\ref{fig:intro_pp}(b), and are approximated by a double exponential form: $\taud(Re) \sim \exp (\exp (a_d Re + b_d))$ and similarly for $\taus(Re)$. (Figure~\ref{fig:tau} discussed below contains actual measured mean lifetimes for channel flow.)   
The timescales cross at a critical value $Re_c$. Below $Re_c$ decay events occur more frequently, while above $Re_c$ splitting events occur more frequently. The crossover between these cases is a key mechanism in the onset of sustained turbulence in wall-bounded shear flow. This crossing point is not, however, the focus of the present study. 

The present study focuses instead on two key issues associated with the rare events themselves. The first is the efficient numerical computation of mean lifetimes. In shear flows, $\taud$ and $\taus$ become extremely large near $Re_c$, making brute force Monte Carlo estimation of mean times exceedingly expensive.
Hence we turn to a more sophisticated class of algorithms that sample rare events by advancing ensembles of trajectories, removing (pruning) unfavourable and duplicating (cloning) favourable ones. 
In particular, we will employ the Adaptative Multilevel Splitting (AMS) algorithm proposed by Cérou \& Guyader \cite{cerou2007adaptive,cerou2011multiple,cerou2019adaptive}. (This nomenclature of "splitting" in the algorithm is unrelated to the splitting of turbulent bands.)
This algorithm impressively paved the way for quantitative study of low-dimensional stochastic systems, as pioneered by Rolland \& Simonnet \cite{rolland2015statistical}, Rolland, Bouchet \& Simonnet \cite{rolland2016computing} or Lestang \emph{et al.} \cite{lestang2018computing}. It was recently applied to large-dimensional fluid-dynamical systems such as atmospheric dynamics \cite{bouchet2019rare, simonnet2021multistability} and bluff-body flow \cite{lestang2020numerical}. 
Rolland \cite{rolland2018extremely} extended the application of this rare-event technique to transitional turbulence, first for transition in a stochastic reduced-order model \cite{barkley2016theoretical} of pipe flow, and then for the collapse of homogeneous turbulence in plane Couette flow \cite{rolland_2022}. 

The second main focus of our study is the origin of the super-exponential dependence of mean lifetimes on Reynolds number, and in particular the connection to extreme values of fluctuations within the one-band state. Goldenfeld, Gutenberg \& Gioia \cite{goldenfeld2010extreme} proposed a mechanism to account for the super-exponential dependence of decay lifetimes of Reynolds number. The essential insight is that the decay process is governed by extreme values and that a linear variation of Reynolds number translates via extreme value distributions to a super-exponential variation in lifetimes. This mechanism was investigated and refined by Nemoto \& Alexakis \cite{nemoto2018method,nemoto2021extreme} in a numerical study of decay events in pipe flow. We will follow a similar analysis applied to both decay and splitting events in channel flow. Finally, the possible connection to the large deviation framework is considered through the computation of most-probable pathways and mean reactive times for rare events. 


\section{Methods}
\label{sec:Methods}
We will now describe two very different types of methods, first, those we use for solving the Navier-Stokes equations governing channel flow, and second, our implementation of the AMS algorithm for capturing rare events.

\begin{figure}[t]
    \centering
 \includegraphics[width=0.65\columnwidth]{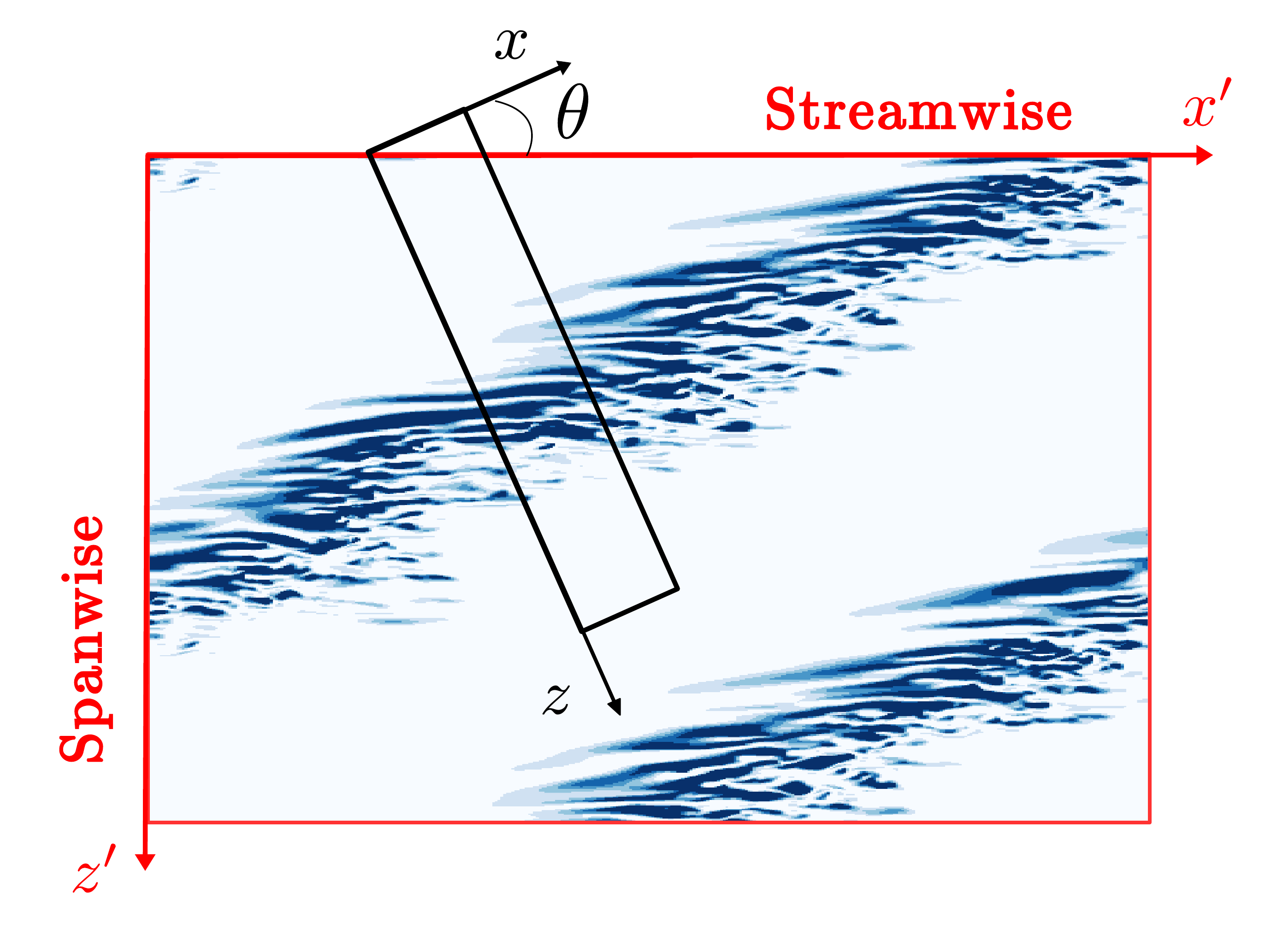} 
    \caption{Visualisation of a turbulent band in a domain periodic in the streamwise and spanwise directions (red bounding box) at $Re=1000$. Colors show transverse energy $\frac{1}{2}(v^2+w'^2)$ in the plane $y=0.8$, from our numerical simulation in a box of size $L_{x'} = 200$, $L_{z'}= 120$. Illustration of the associated tilted computational domain (black) at angle $\theta = 24^\circ$.
  }
\label{fig:domain}
\end{figure}


\subsection{Integration of Navier-Stokes equations in a transitional flow unit}
\label{sec:NS}

The turbulent bands that are the subject of our study are illustrated in figure \ref{fig:domain}. 
We impose a mean velocity $U_\text{bulk}$ on the flow between the two parallel rigid plates. Lengths are nondimensionalised by the half-gap $h$ between the plates, velocities by $3U_\text{bulk}/2$ (which is the centerline velocity of the parabolic laminar flow with mean velocity $U_\text{bulk}$), and time by the ratio between them. The Reynolds number is  defined to be $Re= 3 U_\text{bulk} h/(2\nu)$. 
The non-dimensionalized equations that we simulate are the incompressible Navier-Stokes equations 
\begin{subequations}
    \begin{align}
    \frac{\partial \mathbf{U}}{\partial t}
    + \left(\mathbf{U} \cdot \nabla\right) \mathbf{U} 
    &= -\nabla p  + \frac{1}{Re} \nabla^2 \mathbf{U} \\
    \nabla \cdot \mathbf{U} &= 0
    \end{align}
    \label{eq:NS}
\end{subequations}

Since the bands are found to be oriented obliquely with respect to the streamwise direction, we use a periodic numerical domain which is tilted with respect to the streamwise direction of the flow, shown as the black rectangle in figure \ref{fig:domain}. 
This is common in studying turbulent bands \cite{barkley2005computational,tuckerman2020patterns} and more specifically those in transitional plane channel flow \cite{tuckerman2014turbulent, gome2020statistical,paranjape2020oblique}.
The $x$ direction is chosen to be aligned with a typical turbulent band and the $z$ coordinate to be orthogonal to the band.  The relationship between streamwise-spanwise coordinates $(x',z')$ and tilted band-oriented $(x,z)$ coordinates is:
\begin{subequations}
\label{tilted}
\begin{align}
\mathbf{e}_{x^\prime} &= \quad\cos{\theta} \, \mathbf{e}_x + \sin{\theta} \, \mathbf{e}_z \\
\mathbf{e}_{z^\prime} &= -\sin{\theta } \, \mathbf{e}_x + \cos{\theta} \, \mathbf{e}_z \quad 
\end{align}
\end{subequations}
The usual wall-normal coordinate is denoted by $y$. The field visualised in figure \ref{fig:domain} comes from an additional simulation we carried out in a domain of size ($L_{x^\prime},L_y,L_{z^\prime}) = (200,2,120)$ aligned with the streamwise-spanwise coordinates.

Equations \eqref{eq:NS} are completed by 
rigid boundary conditions in $y$, periodic boundary conditions in $x$ and $z$, and imposed flux 2/3 in the streamwise direction $x^\prime$ and zero in the spanwise direction $z^\prime$:
\begin{subequations}
\begin{align}
&\mathbf{U}(x+L_x,y,z) =
\mathbf{U}(x,y,z+L_z) = \mathbf{U}(x,y,z) \qquad 
\mathbf{U}(x,\pm 1,z) = 0 \\
&\frac{1}{2}\int_{-1}^{+1} dy \: \mathbf{U}(x,y,z) = \frac{2}{3}\mathbf{e}_{x^\prime}=\frac{2}{3} \left(\cos{\theta} \, \mathbf{e}_x + \sin{\theta} \, \mathbf{e}_z\right) \label{eq:flux}
\end{align}
\label{eq:BC}
\end{subequations}
To integrate \eqref{eq:NS} with boundary conditions \eqref{eq:BC}, we use the parallelised pseudospectral C++ code  ChannelFlow \cite{channeflow}, which employs a Fourier-Chebychev spatial discretisation. The velocity field can be decomposed into the stationary laminar parabolic base flow $\mathbf{U_{\text{base}}}=(1-y^2)\mathbf{e}_{x^\prime}$ and the deviation $\mathbf{u}\equiv\mathbf{U} - \mathbf{U_{\text{base}}}$ which satisfies the same equations and boundary conditions as $\mathbf{U}$ but with zero flux instead of \eqref{eq:flux}. A Green's function method is used to impose the flux in each direction. More specifically, for each periodic direction, one computes and uses the pressure gradient such that the resulting flow field will have the desired bulk velocity,  e.g.~\cite{pugh1988two,barkley1990theory}. Throughout our study, we present the deviation $\mathbf{u}= (u,v,w)$ so as to highlight the difference with the dominant laminar flow $\mathbf{U}_{\rm base}$ and the motion of flow features with respect to the bulk velocity.

The angle in this study is fixed at $\theta=24^{\circ}$, as has been used extensively in the past \cite{barkley2005computational,tuckerman2014turbulent,gome2020statistical}. The orientation of the domain imposes a fixed angle on turbulent bands, and choosing a short length for the $x$ direction of the domain suppresses any large-scale variation along the bands. Thus, these simulations effectively capture the dynamics of infinitely long bands that only interact along their perpendicular direction, preventing complex 2D interactions that are possible for finite-length bands \cite{ShimizuPRF2019, xiao2020growth}.
In this way, localised bands in the tilted channel geometry are similar to localised puffs in pipe flow.

Our domain $\domain$ has dimensions ($L_x,L_y,L_z) = (6.6,2,100$) and a numerical resolution of $(N_x,N_y,N_z)=(84,64,1250)$, exactly as in  \cite{gome2020statistical}, thus allowing direct comparison with these prior results. The length $L_z=100$ of our tilted domain corresponds to an inter-band distance above which a band is considered as isolated, while the domain width $L_x=6.6$ is used because it corresponds to the natural spacing of streaks in channel flow in a $24^\circ$ box \cite{chantry2016turbulent,gome2020statistical}.
For puffs in pipe flow, which are similar in many respects to the isolated bands considered here, Nemoto \& Alexakis \cite{nemoto2021extreme} conducted extensive computations showing that domain length had some effect on mean decay timescales, with $L=50$ and $L=100$ giving quantitatively different, but qualitatively similar results. Domain length is expected to have a quantitative effect on the splitting timescale; our domain length $L_z=100$ has been selected as a compromise between accuracy and computational cost.

A semi-implicit time-stepping scheme is used to progress from $\mathbf{u}(t)$ to $\mathbf{u}(u+dt)$, with time step $dt = 1/32=0.03125$. 
Trajectories and associated quantities such as turbulence fraction are sampled at time intervals $\delta t = 32 dt=1$. This sampling time is used throughout for collecting statistics and generating probability distributions.
The computation of solutions of the Navier-Stokes equations discretised in space and time is called, as usual, direct numerical simulation or DNS.


\subsection{The Adaptive Multilevel Splitting (AMS) algorithm}

Here we present the essence of the AMS algorithm. 
We follow closely the method originally described in C\'erou {\em et al.} \cite{cerou2007adaptive}, although here we consider a deterministic dynamical system, the Navier-Stokes equations \eqref{eq:NS}, whereas C\'erou {\em et al.} considered a stochastic process. 
The AMS algorithm has been applied recently to other deterministic fluid systems \cite{lestang2018computing, lestang2020numerical, rolland_2022}. For the application of other rare-event algorithms  to deterministic systems, see \cite{wouters2016rare} and references therein.

\subsubsection{Setup}

Let $\setA$ and $\setB$ be two states visited by trajectories of a dynamical system. More precisely, $\setA$ and $\setB$ are regions in phase space corresponding to particular flow states of interest. We commonly refer to $\setA$ and $\setB$ simply as states. The goal is to produce a large sample of the rare transitions from $\setA$ to $\setB$. In our case $\setA$ will always be the one-band state, labelled as $\setAone$ in Figure~\ref{fig:intro_pp}, while $\setB$ will be either the laminar flow, labelled as $\setBo$, or else 
the two-band state, labelled as $\setBtwo$ in figure~\ref{fig:intro_pp}. 

\begin{table}
    \begin{tabular}{ll}
    \hline
    Symbol & Definition \\
    \hline
$\hA$ & Hypersurface within $\setA$, origin of trajectories, in practice one-band state\\
$\hS$ & Hypersurface $\surfS$ close to and surrounding $\mathcal{A}$\\
$\hB$ &  Hypersurface within $\setB$, destination of trajectories \\
$\hBo$ & Threshold for decay events in AMS \\
$\hBtwo$ & Threshold for splitting events in AMS\\
$\hone$ & Entrance of the \emph{collapse zone} for decays for all $Re$\\
$h_2$ & Entrance of the \emph{collapse zone} for splits for all $Re$\\
$h_M$ & Maximal value of $F_t$ at fixed $Re$\\
$\hleft$ & left endpoint of fit between PDF of $F_t$ and Fisher-Tippett distribution\\
$\hright$ & right endpoint of fit between PDF of $F_t$ and Fisher-Tippett distribution\\
\hline
\end{tabular}
\caption{Definitions of designated levels of a turbulent fraction or score function used throughout the paper.}
\label{tab:hmeaning}
\vspace*{0.3cm}
    \begin{tabular}{l|lllll|lllll}
    \hline
    $Re$ & 815 & 830 & 870 & 900 & 950 & 1000 & 1050 & 1100 & 1150 & 1200 \\\hline
$\hA$ & 0.21 & 0.22 & 0.24 & 0.26 & 0.31 & 0.34 & 0.37 & 0.40 & 0.43 & 0.44\\
$\hS$ & 0.17 & 0.18 & 0.21 & 0.23 & 0.27 & 0.375 & 0.41 & 0.44 & 0.46 & 0.47\\
$\hBo,\hBtwo$ & 0.0001 & 0.0001 & 0.0001 & 0.0001 & 0.0001 & 0.70 & 0.70 & 0.70 & 0.70 & 0.70\\
$\hone,h_2$ & 0.22 & 0.22 & 0.22 & 0.22 & 0.22 & 0.42 & 0.431 & 0.461 & 0.474 & 0.483\\
$h_M$ & 0.292 & 0.305& 0.344 & 0.385 & 0.44& 0.635 & 0.616& 0.659 & 0.677 & 0.69\\
$\hleft$ & 0.13 & 0.148 & 0.176 & 0.207 & 0.243 & 0.30 & 0.32 & 0.279 & 0.271 & 0.326 \\
$\hright$ & 0.285 & 0.278 & 0.307 & 0.327 & 0.364 & 0.42 & 0.436 & 0.469 & 0.501 & 0.536 \\ \hline
\end{tabular}
\caption{Values of designated levels of a turbulent fraction or score function used throughout the paper.}
\label{tab:hRe}
\end{table}

\begin{figure}
    \centering
 \includegraphics[width=\columnwidth]{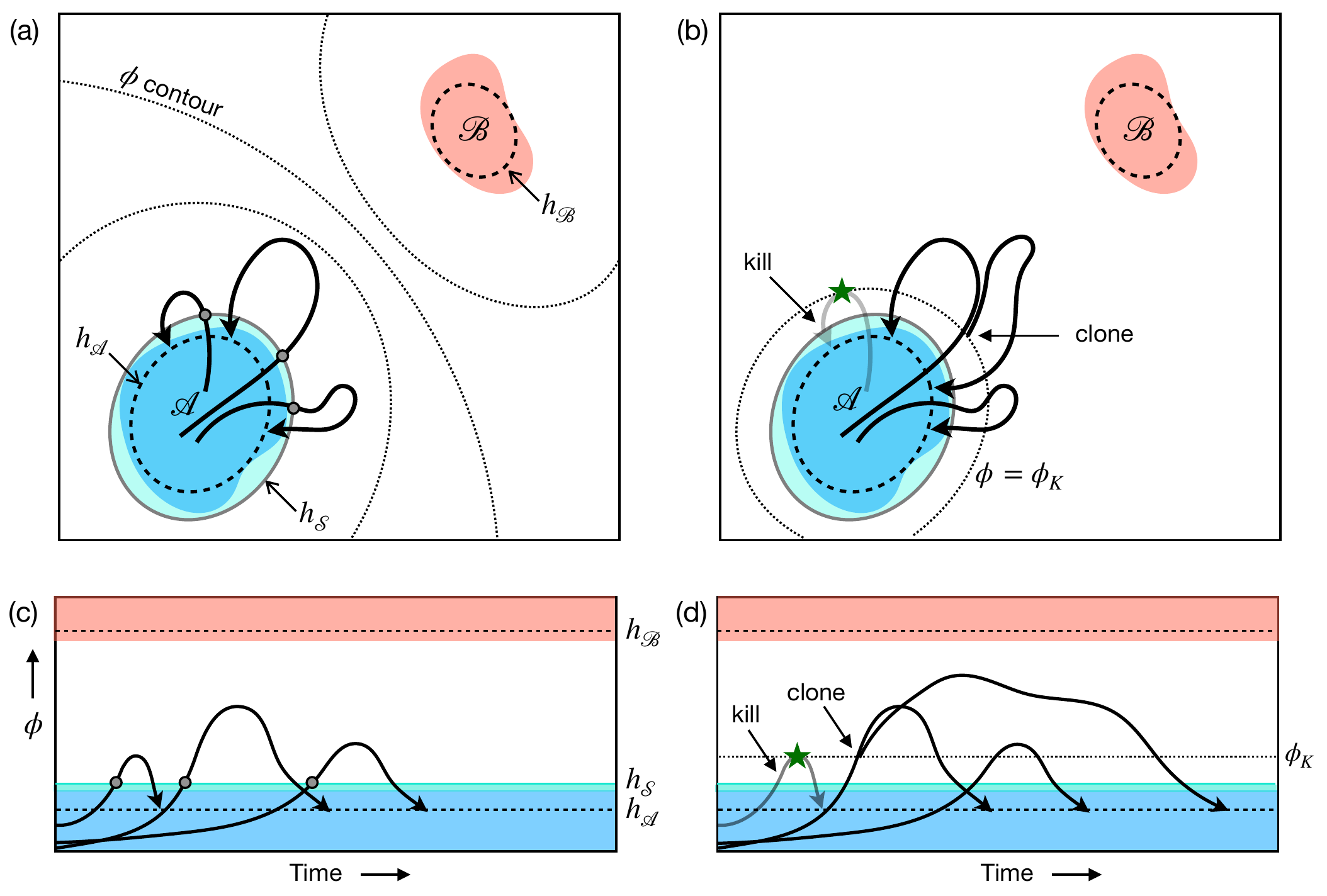} 
    \caption{Schematic depiction of the AMS algorithm for a transition from $\setA$ to $\setB$. (a) The initialisation of the algorithm. Contours are shown for the score function $\phi(\mathbf{u})$ and a hypersurface $\surfS$ surrounding $\setA$. $N$ trajectories are computed starting from random initial conditions in $\setA$ that cross $\surfS$ and then either return to $\setA$ or go to $\setB$. (Here $N=3$ and no initial trajectories reach $\setB$.)  (b) First iteration of the algorithm. The trajectory attaining the smallest maximum score function (here $\phi_{K}$ with $K=1$) is killed, and a new trajectory is cloned from another randomly selected trajectory, resulting in an improved set of trajectories. The process is then iterated until a sufficient number of trajectories reach $\setB$. Time series (c) and (d) correspond to the trajectories in (a) and (b). 
  }
\label{fig:AMS}
\end{figure}

Perhaps the most crucial piece of the AMS algorithm is the specification of a score function, or reaction coordinate, $\phi$, that quantifies transitions from $\setA$ to $\setB$. The score function $\phi(\mathbf{u})$ is a real-valued function of the flow field whose gradient is non-zero (at least everywhere of interest), and such that there exist real values $\hA$ and $\hB$, with $\hA < \hB$, such that $\phi(\mathbf{u}) < \hA$ implies $\mathbf{u} \in \setA$  while $\phi(\mathbf{u}) > \hB$ implies $\mathbf{u} \in \setB$. Note that for decay, the laminar state is a single point in phase space, so we will take $\setB$ to be a set within its basin of attraction. Tables \ref{tab:hmeaning} and \ref{tab:hRe} list the various thresholds of the score function that we will use throughout the paper. The score function provides a smooth landscape for quantifying the progress of the transition between $\setA$ and $\setB$, as illustrated in figure~\ref{fig:AMS}(a). 
The algorithm also requires a value $\hS$ and associated hypersurface $\surfS$, close to $\setA$, given by
$$
\surfS = \{ \mathbf{u} ~| ~\phi(\mathbf{u}) = \hS \}.
$$

\subsubsection{Initialisation}

The initialisation step consists of generating a sample of $N$ trajectories $\mathbf{u}_i(t)$, $i \in \{1, ~ ... ~, ~N\}$, that start within $\setA$, leave $\setA$ at least as far as $\surfS$, and then either reach $\setB$ or, more likely, return to $\setA$. See figure~\ref{fig:AMS}(a).
In practice the $N$ initial conditions $\mathbf{u}_i(0)$ are obtained by taking $N$ snapshots, equally spaced in time, from a single trajectory that remains in $\setA$ over a long time and thus samples the natural measure of states within $\setA$.

The role of the hypersurface $\surfS$ is to ensure that after initialisation, all trajectories in our sample have ventured from $\setA$ at least as far as $\surfS$. Hence the maximum value of the score function obtained along each trajectory is at least $\hS$. From the point of view of the score function, all trajectories in our initial sample have made some, possibly small, progress towards $\setB$. Since $\surfS$ is chosen close to $\setA$, the initialisation step is not computationally demanding. 

For the initialisation and subsequent iterations, it is necessary to store the trajectories. In practice we store full flow fields $\mathbf{u}_i(t_j)$ for each trajectory $i \in \{1, ~ ... ~, ~N\}$ at sparsely spaced times $t_j = j~dT$, as a compromise between the large CPU times required for computing trajectories and the large memory needed to store them. The computations reported here all use a storage interval of $dT = 320~dt=10$, which is 10 times the sampling time $\delta t$ used to collect statistics on trajectories. 

\subsubsection{Iteration}

Iterative step $m$ consists of discarding the $K$ worst-performing trajectories and replacing them with trajectories obtained by cloning non-discarded trajectories. Specifically, we compute the maximal value $\phi^{(m)}_{i}$ of the score function along each trajectory and re-order the trajectories such that 
$$\phi^{(m)}_{1} \leq \phi^{(m)}_{2} \leq \cdots \leq \phi^{(m)}_{K} \leq \cdots \leq \phi^{(m)}_{N}.$$ 
We discard the $K$ trajectories whose maximal values are lowest, in practice a value $K^{(m)} \geq K$ because of possible equality of the maxima. Thus, in general we retain trajectories $\mathbf{u}_i$ such that $\phi_i^{(m)} > \phi_K^{(m)}$.
We replace each discarded trajectory $\mathbf{u}_k(t)$ with a new trajectory constructed as follows:
\begin{enumerate}
\item 
Choose at random (uniformly) one of the trajectories $\mathbf{u}_l(t)$ from the set of $N-K^{(m)}$ retained trajectories. 
Overwrite the trajectory $\mathbf{u}_{k}(t)$ with the part of the trajectory $\mathbf{u}_{l}(t)$ up to time $t^\text{clone}$ at which the score function along $\mathbf{u}_{l}(t)$ first reaches $\phi_K^{(m)}$, i.e.~ $\phi(\mathbf{u}_l(t^\text{clone})) = \phi_K^{(m)}$. See figure~\ref{fig:AMS}(b). (Due to the discrete sampling of stored trajectories, in practice we copy trajectories until the score function first exceeds $\phi_K^{(m)}$.)

\item Modify $\mathbf{u}_{l}(t^\text{clone})$ with a low-amplitude multiplicative spectral perturbation as follows. Let
$$\bm{\eta}(x,y,z) = \sum_{m_x} \sum_{m_z} \sum_{m_y} \mathbf{\tilde{\eta}}_{m_x,m_y,m_z} s^{|m_x|+|m_y|+|m_z|} e^{i(m_x k_x x + m_z k_z z)} T_{m_y}(y)$$
where each $\mathbf{\tilde{\eta}}_{m_x,m_y,m_z}$ is a vector whose components are uniform random complex numbers 
of modulus less than 1, 
$s$ is a smoothing parameter such that $0<s<1$, 
and $T_{m_y}$ is the Chebyshev polynomial of order $m_y$.
Then the low-amplitude multiplicative perturbation at the cloning time is
\begin{equation}
\mathbf{u}_k(x,y,z,t^\text{clone}) =  (\bm{I} + \epsilon \bm{\eta}(x,y,z))  \mathbf{u}_l(x,y,z,t^\text{clone})
\label{eq:pert}
\end{equation}
where $\epsilon$ sets the size of the perturbation.
The weak random perturbation is necessary to ensure that cloned trajectories do not exactly repeat the path of the trajectory from which they are cloned. Perturbations are always sufficiently weak that they leave the score function unchanged to at least four significant digits.
Rolland~\cite{rolland_2022} uses a similar approach in applying AMS to turbulence collapse in Couette flow. The remainder of the trajectory $\mathbf{u}_{k}(t)$ for $t>t^\text{clone}$ is obtained by simulating the new trajectory until it reaches $\setA$ or $\setB$ as before. 

\end{enumerate}

Once the $K^{(m)}$ discarded trajectories have been replaced (overwritten), we have a new set of $N$ trajectories that are superior to the set at the start of the iteration, in the sense of being closer to reaching $\mathcal{B}$. Specifically, the maximum value of the score function for each of the new trajectories is now at least $\phi^{(m)}_K$. We increment $m$ and repeat as necessary. 

\subsubsection{Stopping and post processing} 

Iterations end once the $N$ samples have all reached $\setB$. The final number of iterations is denoted by $M$. From the resulting trajectories and information gathered during the iteration process, we can construct estimators of relevant statistical quantities. 
Trajectories begin in $\setA$, pass through $\surfS$ and terminate upon arrival at either $\setA$ or $\setB$. 
The estimator of the probability to go from $\surfS$ to $\setB$ is given by \cite{cerou2007adaptive}:
\begin{equation}
    \hat{p} = \prod_{m=1}^{M} \left(1 - \frac{K^{(m)}}{N}\right), 
    \label{eq:pams}
\end{equation}
where $K^{(m)}$ is the number of trajectories eliminated at iteration $m$. The probability of going from $\surfS$ to $\setA$ is $(1-\hat{p})$ and that of going from $\setA$ to $\surfS$ is 1.

\begin{figure}[t]
    \centering
 \includegraphics[height=3cm]{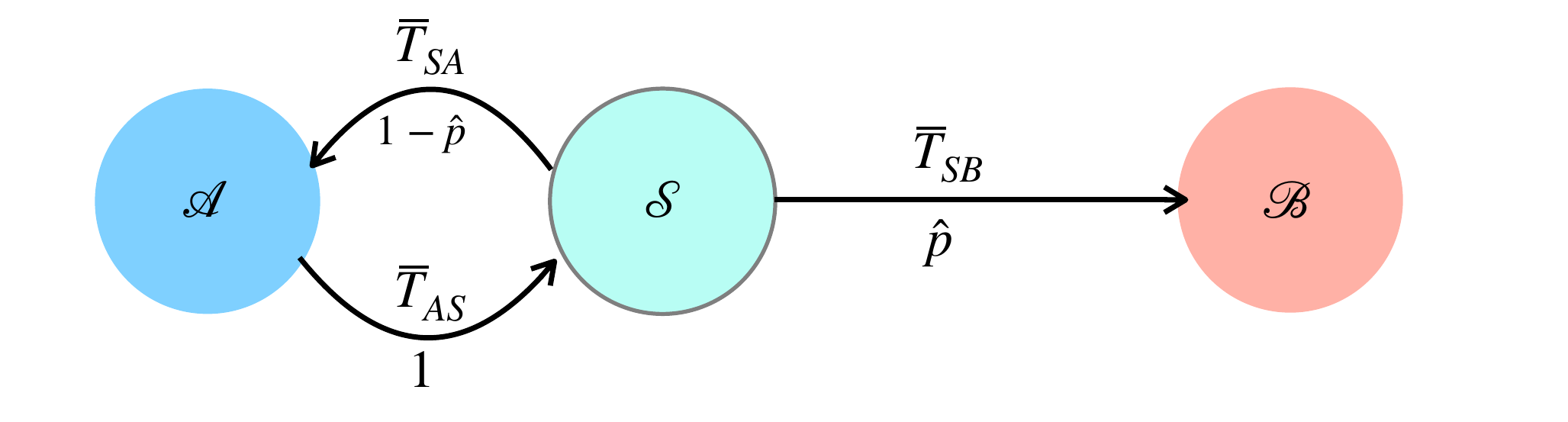} 
    \caption{Schematic depiction of the data gathered via the AMS algorithm for a transition from $\setA$ to $\setB$ via $\surfS$. The probability $\hat{p}$ of transition from $\surfS$ to $\setB$ is estimated, thus giving $1-\hat{p}$ as the probability for transition from $\surfS$ to $\setA$. The sample mean times obtained for the two transitions are $\oT_{\surfS\setB}$ and $\oT_{\surfS\setA}$. From $\setA$, all trajectories reach $\surfS$ (probability of one) and the sample mean time for this transition is $\oT_{\setA\surfS}$. Trajectories begin at $\setA$ and make some number of round trips between $\surfS$ and $\setA$ before possibly reaching $\setB$.
  }
\label{fig:Schematic_Markov}
\end{figure}
The main quantity of interest is the mean first passage time $\tau$ from state $\setA$ to state $\setB$. 
For this, we will require the sample mean times available from the computations~\cite{cerou2011multiple}. Let $T_{\setA\surfS} \equiv \inf \{ t>0, ~ \mathbf{u}(t) \in \surfS ~|~ \mathbf{u}(0) \in \setA \}$ and let $\oT_{\setA\surfS}$ denote its sample mean 
obtained from trajectories whose initial conditions $\mathbf{u}(0)$ are selected
from a long simulation lying within $\setA$. 
Because $\surfS$ is close to $\setA$,  $\oT_{\setA\surfS}$ is easily obtained from DNS (or from the initialisation step of the AMS). Similarly, from the trajectories that cross $\surfS$ and return to $\setA$ we can compute $\oT_{\surfS\setA}$, the sample mean time to go from $\surfS$ to $\setA$. Finally, from the $N$ sample paths constructed as part of the AMS we can compute $\oT_{\surfS\setB}$, the sample mean time to go from $\surfS$ to $\setB$. 

From these quantities, the estimator for the mean first passage time $\tau$ is constructed as illustrated in  figure~\ref{fig:Schematic_Markov}.
A trajectory going from $\setA$ to $\setB$ does so by going from $\setA$ to $\surfS$ and back some number of times, $n$, before finally transitioning from $\setA$ to $\surfS$ to $\setB$. 
The probability of such a trajectory is $(1-\hat{p})^n~\hat{p}$ and the mean time for all such trajectories is
$
\left( \oT_{\setA\surfS}+ \oT_{\surfS\setA} \right)n+ \oT_{\setA\surfS}+\oT_{\surfS\setB}.
$
Summing over all possible $n$ yields the estimator for $\tau$:
\begin{align}
    \tau &= \sum_{n=0}^\infty (1-\hat{p})^n\hat{p} \left[\left( \oT_{\setA\surfS}+ \oT_{\surfS\setA} \right)n+ \oT_{\setA\surfS}+\oT_{ \surfS\setB}\right]\nonumber\\
&=\left( \oT_{ \setA\surfS}+ \oT_{\surfS  \setA} \right) \frac{1-\hat{p}}{\hat{p}} +
    \left(\oT_{\setA\surfS}+ \oT_{\surfS  \setB}\right). 
    \label{eq:tau}
\end{align}
We do not use separate notation for the true mean first passage time and this estimator of it. 
In describing the transition dynamics in terms of a Markov chain in figure~\ref{fig:Schematic_Markov}, we rely on standard assumptions of the AMS algorithm, stated by C\'erou \emph{et al.} \cite[p. 12] {cerou2011multiple}.

The time $\oT_{ \setA\surfS}+ \oT_{\surfS  \setA}$ is the mean {\it non-reactive time}. This is the mean time for trajectories starting from within $\setA$ to return to $\setA$, conditioned on the fact that they reach $\surfS$. Similarly, $\oT_{\setA\surfS}+ \oT_{\surfS \setB} $ is the mean {\it reactive time} for trajectories starting from within $\setA$ to reach $\setB$, conditioned on the fact that they do not return to $\setA$. 
Neither the reactive time nor the non-reactive time is particularly large. What makes the mean first passage time large is that on average a trajectory will make many failed attempts to reach $\setB$ so that the mean non-reactive time is multiplied by the large factor $(1-\hat{p})/\hat{p}$.


\section{Computing mean passage times in channel flow}
\label{sec:time_scales}

\subsection{Choice of the score function for band decay and splitting}

The choice of the score function is critical for the AMS algorithm. In our case we need functions that quantify the transition progress between the one-band state $\setAone$ and either the laminar state $\setBo$ (decay event) or the two-band state $\setBtwo$ (splitting event). We use slightly different score functions for decay and splitting.

We introduce the turbulent fraction, $F_t$, quantifying the proportion of the flow that is turbulent: $F_t = 0$ for laminar flow, while $F_t = 1$ for flow that is turbulent throughout the channel. For localised turbulent bands, the turbulent fraction is between zero and one. Specifically we define
\begin{equation}
e(z) \equiv \frac{1}{L_x L_y} \int_{-1}^1 \int_0^{L_x} \frac{1}{2} ( v^2 + {w^\prime}^{2}) ~ \mathrm{d}x \,\mathrm{d}y,
\quad \text{and}\quad
F_t \equiv  \frac{1}{L_z}\int_0^{L_z} H(e(z) - \ethr) \,\mathrm{d}z \label{eq:Ft}
\end{equation}
where $H$ is the Heaviside function. These quantities use the energy contained in the cross-channel and spanwise velocity components $v$ and $w^\prime$, which is zero for laminar flow. Its cross-sectional integral $e(z)$ provides a good characterisation of the turbulence as a function of $z$. We define the flow at location $z$ to be turbulent if $e(z)$ exceeds the empirical threshold $\ethr$, where $\ethr = 1.1 \times 10^{-3}$.
\begin{figure}[t]
   \centering
    \subfloat[]{\includegraphics[width=\columnwidth]{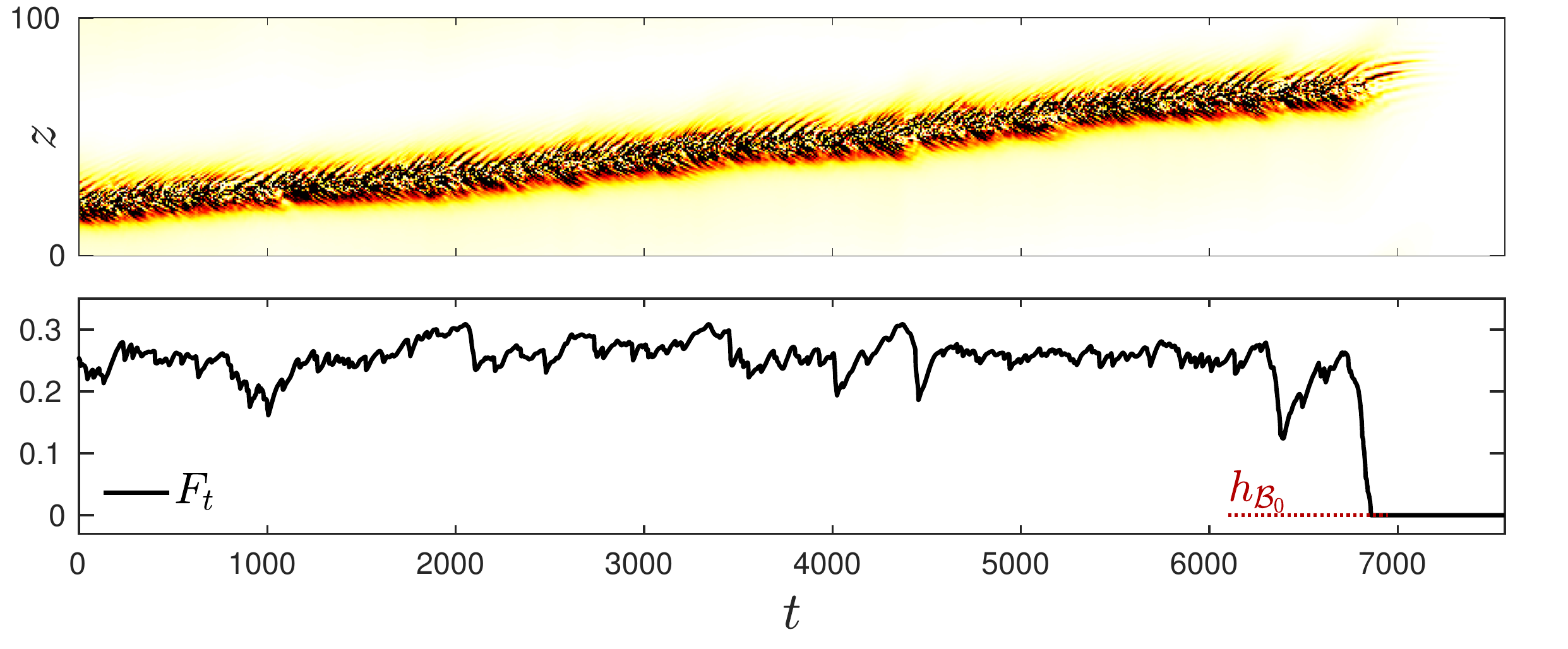}
    \label{fig:score_probe_decay}}
    \vspace*{-10pt}
    \subfloat[]{\includegraphics[width=\columnwidth]{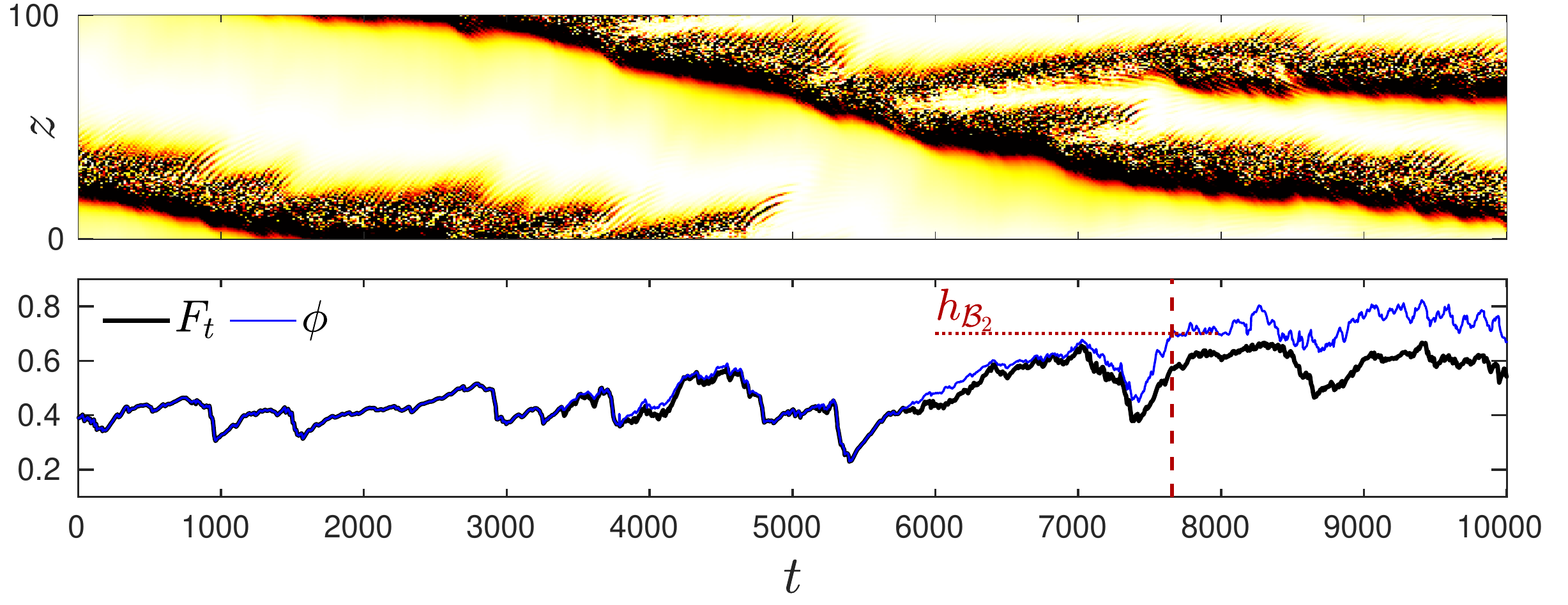} \label{fig:score_probe_split}}
    \caption{Evolution of the turbulent band during (a) a decay at $Re=870$ and (b) a split at $Re=1150$. Top: Spatio-temporal visualisation. Colors show $(v^2 + w'^2)/2$ at $(x=3.3,y=0.8)$ (white: 0, black: 0.001). Bottom: Evolution of the turbulent fraction $F_t$ (black curves) and of score function $\phi$ (thin blue curve) defined for splits in \eqref{eq:score}. 
    }
    \label{fig:score}
\end{figure}
Figure \ref{fig:score_probe_decay} presents the typical life of a decaying band, repeated from figure~\ref{fig:intro_time}, along with the corresponding time series of the turbulent fraction $F_t$. Local minima of $F_t$ occur at local contractions of the band, which are themselves small detours towards the laminar state. Then $F_t$ drops sharply to zero when the band transitions to the laminar state.
In practice, we take $\phi = F_t$ and replace $<$ with $>$ (and max with min) as necessary in the algorithm.
We define the system to be in $\setBo$ if $\phi < \hBo = 0.0001$ independently of $Re$, since all trajectories attaining this value of $F_t$ are in the basin of attraction of the laminar state. The value $\hAone$ is taken as the most probable value of the score function from a long simulation of the one-band state. As a result, $\hAone$ depends on Reynolds number. The level $\hS$ is chosen to be approximately $0.8\,\hAone$. (See also Tables \ref{tab:hmeaning} and \ref{tab:hRe} for definitions and values of all of these levels.)

We now consider the transition from one to two bands. Unlike for band decay, we have found that the turbulent fraction is not an adequate score function for band splitting.  
Figure~\ref{fig:score_probe_split} illustrates the difficulty.
We see that before attaining the two-band state, multiple attempts to split occur. These deviations from the one-band state are characterised by widening of the initial band, possibly leading to the opening of a laminar gap between two turbulent regions. The resulting downstream turbulent patch then either decays, leading to a one-band state, or gains in intensity, ultimately leading to a steady second turbulent band whose shape and energy level are comparable to those of the initial band. 
The problem with using $F_t$ as a score function is that while it captures the widening of the single band, it does not select for the intensification of downstream patches that results in a persistent secondary band. 
In figure~\ref{fig:score_probe_split}, the branching which will eventually lead to a new band occurs at $t\approx 5400$, but it is only at $t\approx 7660$ that this band becomes wider and more intense, acquiring some permanence and stability. It is this latter event that we will define as the split.

We have constructed an empirical but successful score function $\phi$ that encompasses the entire process of band stretching, captured by $F_t$, as well as separation into multiple bands and subsequent intensification of downstream bands. As can be seen by comparing the blue and black curves in  figure~\ref{fig:score_probe_split}, $\phi$ does not differ greatly from $F_t$, but the difference is crucial for the performance of the AMS algorithm. 
The score function is given as follows. Consider the flow to consist of $n_b$ turbulent bands, i.e.~$n_b$ distinct regions in which $e(z)>\ethr$, as defined in \eqref{eq:Ft}.
We associate to each turbulent band its width $W_i$ in $z$, the laminar gap length $L_i$ upstream until the next turbulent band, and finally its average energy $E_i$.
We consider the mother band to be the band whose upstream laminar gap is maximal. Its properties are labeled $(W_{1} , L_1, E_{1}) $, and the other bands $i$ are ordered by downstream distance from the mother band. We then define the following empirical score function for splits:
\begin{equation}
\label{eq:score}
 \phi = F_t + \sum_{i=1}^{n_b} \frac{l_{i} }{L_z} \left(\frac{E_{i}}{E_{\max}}\right)^\alpha  =   \frac{1}{L_z}\sum _{i=1}^{n_b} \left[{W_{i} + l_{i} \left(\frac{E_{i}}{E_{\max}} \right)^\alpha}\right]
\end{equation}
Here, $E_{\max} \equiv \underset{1 \leq i \leq n_b}{\max} {E_i}$ and $l_i \equiv \sum_{j = 2}^i  L_j$ is the total laminar gap between band $i$ and the mother band, which can describe continuously the collapse or splits of multiple child bands.
The exponent $\alpha$ is chosen empirically to balance energy localization and turbulence spreading. In practice, we use $\alpha=0.5$, in order to counteract the decrease in turbulent fraction usually observed after a split, as shown on figure~\ref{fig:score_probe_split} at $t=7\,500$.
In this way, we have enhanced the turbulent fraction by adding a function of band intensity $E_i$ and of the total laminar distance $l_i$ to the mother band.
In this study, the level $\hBtwo = 0.7$ is found to capture a successful split: the presence of a lasting secondary band whose profile and intensity are similar to those of the initial band.
We take $\hS \simeq 1.2 \hAone$, with $\hAone$ the most probable value of \eqref{eq:score} in the one-band state.

We have introduced a number of numerical parameters that could affect the performance and the accuracy of the computations. Of these, the selection of $\hBtwo$ and $\epsilon$ require the most care. Referring to figure~\ref{fig:score_probe_split} one sees that the threshold $\hBtwo$ must correctly capture the completion of a splitting event. As with the difficulty in defining a good score function for splitting, this is a reflection of our lack of good understanding of the splitting process. As can be seen in figure~\ref{fig:score_probe_decay}, this issue does not arise for decay since the score function of the laminar state is known to be zero. Concerning the perturbation size $\epsilon$ used in the cloning, equation \eqref{eq:pert}, one would ideally choose this to be small and independent of $Re$. In practice we have found it necessary to vary $\epsilon$ with $Re$, and as discussed at the end of section \ref{sec:time_scales}\ref{sec:rare_events}, the current algorithm applied to decay events sometimes requires $\epsilon$ to be larger than desired. (See the Supplemental Material for further discussion of the perturbation size $\epsilon$ and also the sample size $N$.)


\subsection{Simulating rare events with AMS}
\label{sec:rare_events}

\begin{figure}[t]
    \centering
    \includegraphics[width=0.7\textwidth]{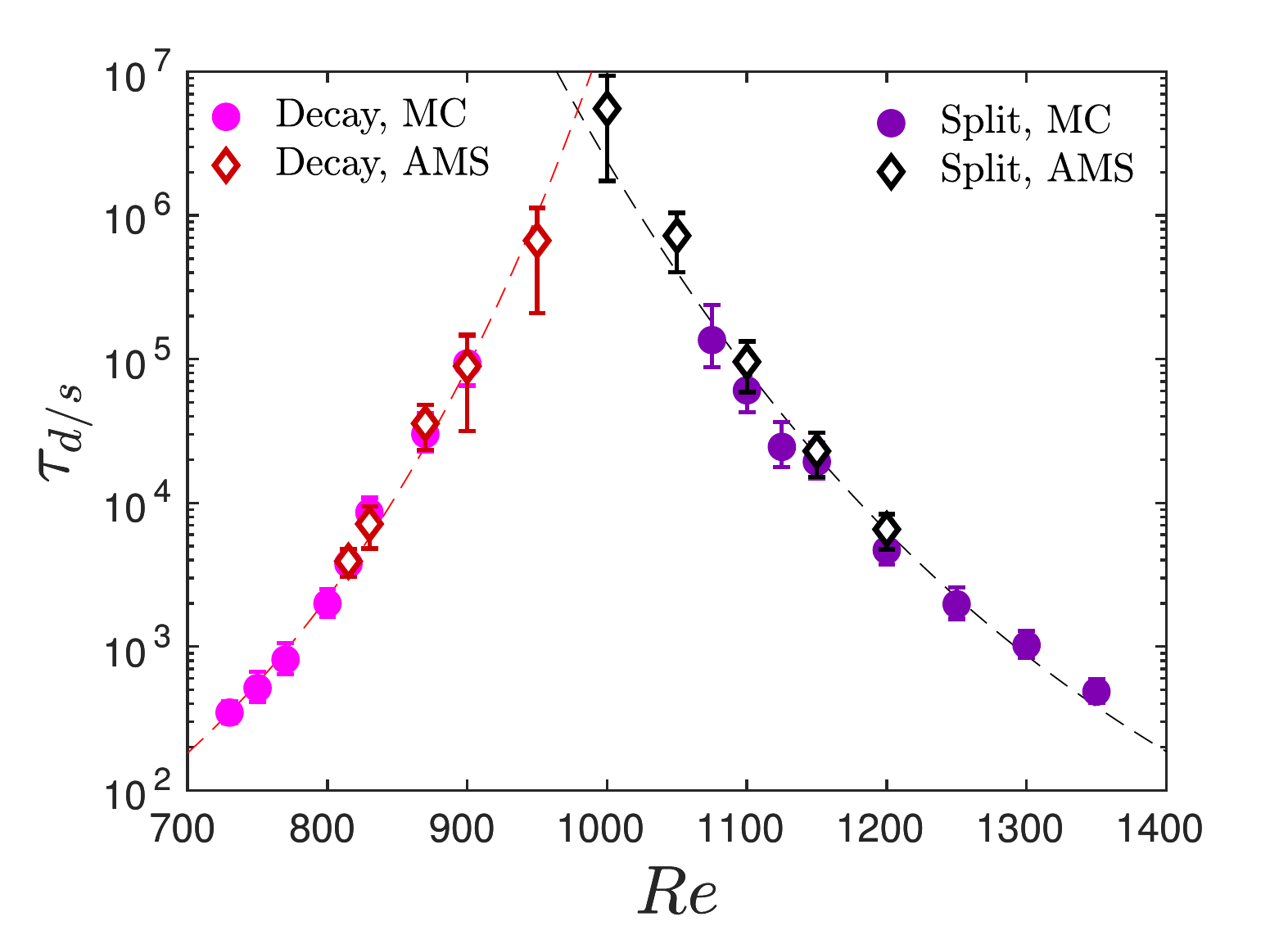}
    \caption{Mean decay times (red, magenta) and splitting times (black, purple) of turbulent bands as a function of Reynolds number, estimated with the Monte Carlo method (MC, circles) or with the Adaptative Multilevel Splitting (AMS, diamonds). Error bars give confidence intervals for MC and are computed from multiple realizations of the algorithm for AMS. Dashed lines are best fits to double exponential form using the combined AMS and MC data: 
    $\tau_d \simeq \exp{[\exp{(3.9\times10^{-3}~Re - 1.09)}]}$; 
    $\tau_s \simeq \exp{[\exp{(-2.6\times10^{-3}~Re +5.27)}]}$.
    }
    \label{fig:tau}
\end{figure}

We have used the AMS algorithm to compute the mean decay and splitting times of an isolated turbulent band in a channel. These mean times are plotted as a function of Reynolds number in Figure \ref{fig:tau}, where we also include previous results obtained via standard Monte Carlo (MC) simulations~\cite{gome2020statistical}. The AMS results overlap with the Monte Carlo data, but also substantially extend the range of accessible time scales. Both the AMS and Monte Carlo results use the same tilted computational domain, the same spatial resolution, and the same underlying time-stepping code, as described in section \ref{sec:Methods}.\ref{sec:NS}. This permits direct comparison of the two methods. 

Figure \ref{fig:tau} confirms the super-exponential dependence of the time scales found for decay and splitting events in wall-bounded shear flows~\cite{avila2010transient,avila2011onset,shi, gome2020statistical}. From fits with $\tau_d = \exp{( \exp (a_d ~ Re+b_d))}$ and $\tau_s = \exp{( \exp (a_s ~ Re+b_s))}$ in the decay and split regimes, we find $Re_c \simeq 980$ as an improved estimate of the crossing Reynolds number for this flow configuration. (Previous fits to the Monte Carlo data gave a crossing Reynolds number of $965$.)

We recall a few details from the Monte Carlo computations in \cite{gome2020statistical}. The initial fields for the simulations are taken from snapshots of long-lasting bands simulated at $Re \in [900 - 1050]$. The Reynolds number is then changed to the desired value. Decay and splitting times from the start of the simulation are recorded. From these, the mean times and associated error bars are obtained \cite{gome2020statistical}.
The Monte Carlo estimate of the transition probability $\pMC$ is computed from the multiple simulations by counting the number of decays or splits relative to the number of passages through $\surfS$.
Typically $N=40$ decay and splitting events are obtained at each Reynolds number. Fewer than $N=40$ events were obtained by Monte Carlo at the largest values of $\tau$. With such techniques, only time scales $\tau < 10^5$ are currently accessible in practice.

The AMS initial fields are created from long-lasting bands, as in the Monte Carlo method, except that each initial field is simulated for an additional relaxation time of $500$ before commencing the AMS algorithm. 
The number of trajectories we seek to discard at each AMS iteration is $K=1$.
At each value of $Re$, the AMS algorithm is run $N_\text{AMS}$ times, with each realisation computing a sample of $N$ trajectories. 
Each realisation gives a value of $\tau$ calculated using \eqref{eq:tau}, where $\overline{T}_{\setA \surfS}+\overline{T}_{\surfS \setA}$ is computed by DNS as part of the initialisation step, $\overline{T}_{\setA\surfS}+\overline{T}_{\surfS\setB}$ is obtained from the AMS trajectories, and $\hat{p}$ is obtained via \eqref{eq:pams}. Then the final estimate of $\tau$ is obtained by averaging over the $N_\text{AMS}$ independent realisations. 

Table \ref{tab:ams_results} compares estimates of the transition probability $\hat{p}$ from the Monte Carlo and AMS strategies. Both methods yield comparable estimates when Monte Carlo results can be obtained. We emphasise that lifetimes $\tau$ change by orders of magnitude over the range of $Re$ of interest, so we do not seek more than about one digit of accuracy in their values.  
The overall gain in computational speed achieved by the AMS over Monte Carlo is measured by the total CPU time. One component of this cost is the CPU time per trajectory, for which the AMS shows a typical improvement of order $\mathcal{O}(10)$ and even $\mathcal{O}(100)$ for the low-transition-probability cases we considered; see $Re=1000$ in Table~\ref{tab:ams_timing}. For higher-transition-probability cases, AMS does not outperform Monte Carlo because AMS requires $N_\text{AMS}$ realisations to compensate for the variability in individual realisations. For low-transition-probability cases such as $Re=1000$, only AMS is capable of inducing the very rare trajectories which are out of reach for the Monte Carlo method. (See Supplemental Material for further comparisons.)

The results from AMS show larger variability than those from Monte Carlo, especially for decay cases, as seen by the error bars on figure~\ref{fig:tau}.
It is known that the standard deviation of the estimated probability for AMS will decrease as $1/\sqrt{N}$ (at least in ideal cases)~\cite{brehier2016unbiasedness,rolland2015statistical}. For our high-dimensional system, $N$ is restricted by computational costs. Using $N$ larger than $100$ is not practical and we typically use $N = 50$.
We observe that the large variability between different realisations of the AMS algorithm is associated with variability in the initialisation, especially the extent to which the initial trajectories are a representative sample.

\begin{table}
    \centering
    \resizebox{\textwidth}{!}{
    \begin{tabular}{|c||c|c|c||c|c|c|c|}
        \hline
         $Re$ & \multicolumn{3}{c||}{Monte Carlo (MC)} & \multicolumn{4}{c|}{Adaptive Multilevel Splitting (AMS)}  \\
         \cline{2-8}
          & $N$  & $\pMC$ & $\tau_\text{MC}$ &
          $\epsilon$ & $N_\text{AMS} \times N$  & $\pAMS$ & $\tau$ \\
         \hline
         870~~ & 40 & 0.081 &$3.0 \times 10^4$
         & $5 \times 10^{-4}$ &$9 \times 50$ & 0.081~~~~ & $3.6 \times 10^4$ \\
         \hline
         900~~ & 40 & $0.013$ & $9.3 \times 10^4$ &
         $1 \times 10^{-3} $& $7\times50$ & 0.015~~~~ & $8.9 \times 10^4$\\
         \hline
         1000 & -- & -- & -- & $1\times 10^{-3}$ &$3\times50$ &$0.00029$&$5.5\times10^6$ \\
         \hline
         1150~~~ & 40 & 0.047 &$2.1\times10^4$ &
         $1\times10^{-5}$& $9\times50$ & 0.046~~~~ & $2.2\times10^4$
         \\
         \hline
    \end{tabular}}
    \caption{Results of Monte Carlo (MC) and AMS (Adaptative Multilevel Splitting). $N$ is the number of samples for MC or for a single realisation of AMS. For AMS, $N_\text{AMS}$ is the number of realisations of the algorithm and $\epsilon$ is the perturbation amplitude used in cloning. The estimated transition probability and mean first passage time obtained by MC and AMS are $\pMC$, $\tau_\text{MC}$ and $\hat{p}$, $\tau$, respectively.}
    \label{tab:ams_results}
    \vspace*{0.5cm}
    \centering
    \resizebox{\textwidth}{!}{
    \begin{tabular}{|c||c|c|c||c|c|c|c|}
        \hline
         $Re$ & \multicolumn{3}{c||}{Monte Carlo (MC)} & \multicolumn{4}{c|}{Adaptive Multilevel Splitting (AMS)}  \\
         \cline{2-8}
          & $N$  & CPU$_{\text{traj}}$ & CPU$_\text{tot}$  &
          $\epsilon$ & $N_\text{AMS} \times N$ & CPU$_\text{traj}$ & CPU$_\text{tot}$ \\
         \hline
         870~~ & 40 & 2500 & $1 \times 10^5$  
         & $5 \times 10^{-4}$ &$9 \times 50$ & 360 &$1.6\times10^5$ \\
         \hline
         900~~ & 40 & 7500& $3 \times 10^5$  &
         $1 \times 10^{-3} $& $7\times50$ &330 & $1.2\times10^5$ \\
         \hline
         1000$^{*}$ & 40 & $4\times10^5$ & $2\times10^7$  & $1\times 10^{-3}$ &$3\times50$ &1000& $1.5\times10^5$ \\
         \hline
         1150~~~ & 40 & 5000 & $2\times10^5$ &
         $1\times10^{-5}$& $9\times50$ & 500& $2.2\times10^5$ 
         \\
         \hline
    \end{tabular}}
    \caption{Performance of Monte Carlo (MC) and AMS (Adaptative Multilevel Splitting). $N$ is the number of samples for MC or for a single realisation of AMS. For AMS, $N_\text{AMS}$ is the number of realisations of the algorithm and $\epsilon$ is the perturbation amplitude used in cloning. 
    The estimated CPU time per successful trajectory is given, as well as the total CPU time (both in processor hours on an HPE SGI 8600 computer). $^*$For $Re=1000$, no estimator of the time scale could be achieved by Monte Carlo, so the CPU times are extrapolated from the AMS estimator $\tau$.}
    \label{tab:ams_timing}
    \vspace*{-4pt}
\end{table}

The amplitude $\epsilon$ of the perturbation that we use in cloning trajectories is chosen to promote separation of the trajectories. The only issue occurs for rare decay ($Re \in [900 - 950]$) where the amplitude must be increased ($\epsilon > 10^{-2}$ at $Re=950$). 
In these cases, cloned trajectories resulting from small-amplitude perturbations separate from one another only after having reached their minimum $F_t$ value. Hence they do not acquire an improved score function, causing the algorithm to stagnate.
The reason for this is that the duration of the approach to the minimum of $F_t$ is shorter than the Lyapunov time of the system. 
This limitation of our current procedure has been observed in other studies \cite{lestang2020numerical,rolland_2022} and has been addressed in \cite{rolland_2022} by anticipating branching. This technique clones trajectories prior to where one would in the standard algorithm, thus promoting the separation of trajectories near the minimum of $F_t$.


\section{Extreme value description of decay and splitting trajectories}
\label{sec:extreme}

The super-exponential dependence of lifetime of turbulence on Reynolds numbers seen in figure \ref{fig:tau} is ubiquitous for decay and splitting events in wall-bounded shear flows, e.g. \cite{hof2008repeller,avila2010transient, avila2011onset,shi,gome2020statistical}.
Goldenfeld, Gutenberg \& Gioia \cite{goldenfeld2010extreme} have formulated a hypothesis explaining decays through extreme value theory. The main idea is to associate the decay of a turbulent patch to the statistical distribution of the largest fluctuation over some space-time interval. If the maximum amplitude of fluctuations becomes lower than some threshold, then the multiple fluctuations comprising a turbulent zone will all laminarise. This connects laminarisation to the distribution of extrema of a set of random variables. Just as the Central Limit Theorem states that under very general conditions the limit of the sum of independent and identically distributed random variables is a Gaussian, the Fisher-Tippett-Gnedenko theorem \cite{fisher_tippett_1928} states that the extrema of a set of $n$ independent and identically distributed variables should follow a Fisher-Tippett distribution. Goldenfeld {\em et al.} assumed that the decay threshold depends on $Re$ and approximated that dependence locally as linear. This linear dependence translates into a super-exponential dependence of the lifetimes on $Re$ via properties of the Fisher-Tippett distribution.

In a study of the decay of turbulent puffs in pipe flow, Nemoto \& Alexakis \cite{nemoto2021extreme} found that the maximal vorticity over the domain followed a Fr\'echet distribution, a member of the Fisher-Tippett family. Moreover, they found that the parameters of this distribution depend linearly on $Re$ over a range of $75$ in $Re$ near the critical value $Re_c$. Similar to the Goldenfeld {\em et al.} argument, this linear dependence on parameters translates to a super-exponential dependence of the lifetimes on $Re$. 
Thus, Nemoto \& Alexakis were able to directly relate extreme values to the super-exponential evolution with $Re$ of the puff decay times in pipe flow. 
Other quantities related to the distance to the laminar attractor have been shown to follow the extreme value law \cite{manneville2011decay, shimizu2019exponential}, particularly when a maximal or minimal value is extracted from a divided time series \cite{faranda2014using}. 

Here we explore these ideas for both the decay and splitting of turbulent bands in channel flow over a substantial range of $Re$. To do so, we must link the rare events (decay or split) with some observable that follows an extreme distribution.
Rather than speculate on which variable or combination of variables are mechanistically responsible for driving decay and splitting events, we choose to focus on $F_t$ for both transitions. Our reasoning is that turbulence fraction is a useful observable of general interest that is easily obtainable in computations and experiments. 
As we show below, the turbulent fluctuations and reaction pathways project onto $F_t$ and allow us to analyse the connection between fluctuations and the rare events.
As a practical matter, it is helpful to study distributions of a quantity that is (or is closely related to) the score function used to obtain rare events. 


\subsection{Probability densities of turbulent fraction}

\begin{figure}[h!]
    \centering
    \includegraphics[width=\textwidth]{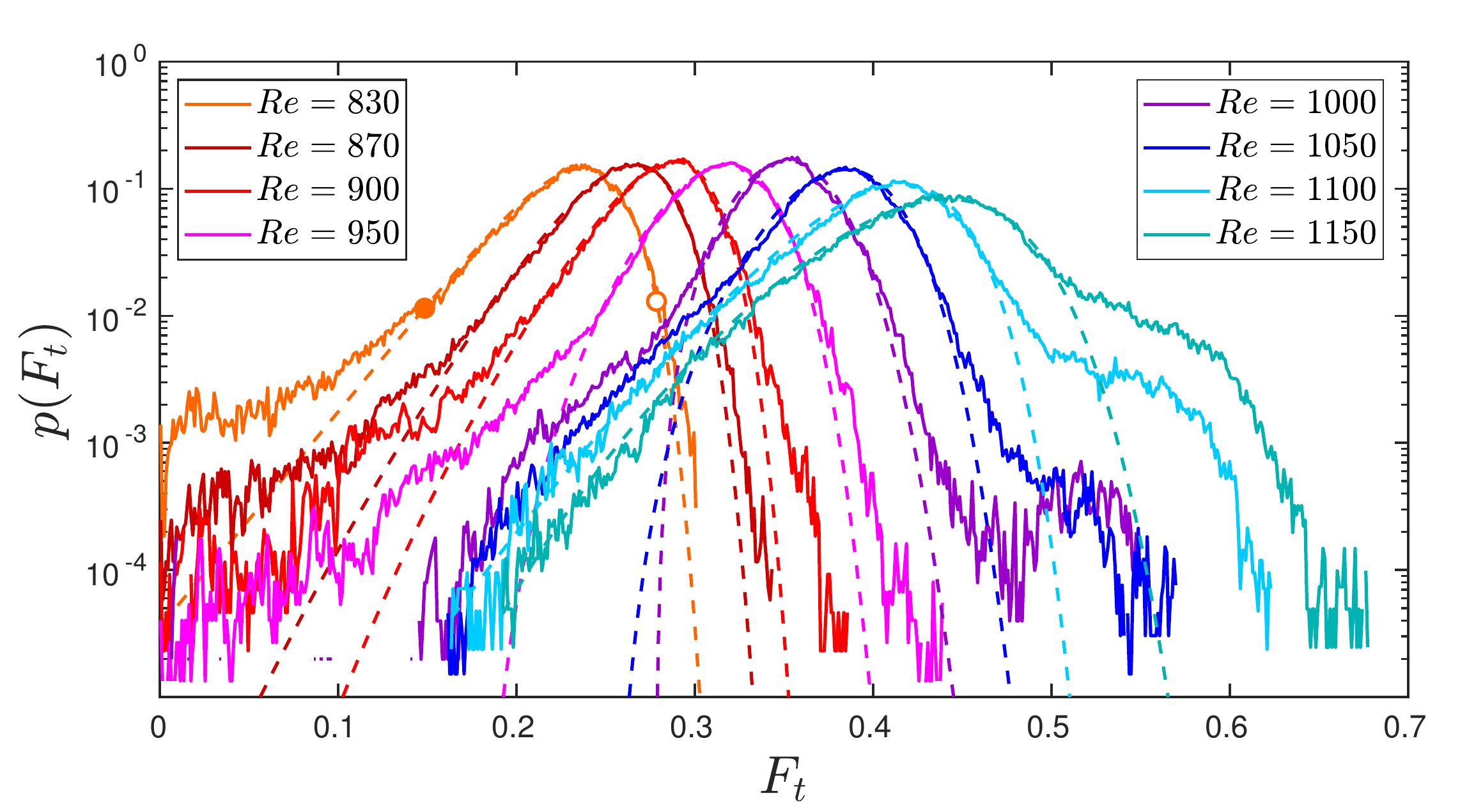}
    \caption{Probability density function of the turbulent fraction around the one-banded state $\setAone$. Dashed lines correspond to fits with a Fisher-Tippett probability density, the derivative of \eqref{eq:FisherTippett}. Fits are carried out over intervals $[\hleft,\hright]$, shown for the case $Re=830$ by colored and white points. Values of $\hleft$ and $\hright$ are given in table \ref{tab:hRe}.
    }
    \label{fig:pdf_fit}
\end{figure}

We begin by showing in figure~\ref{fig:pdf_fit} the probability density function (PDF) of the turbulent fraction $p(F_t)$ for a variety of Reynolds numbers. These PDFs have been constructed from Monte Carlo simulations that start, after initial equilibration time, from the one-band state $\setA$ and terminate at the end of a decay or split. The distributions have a clear asymmetry about their maxima and they have broad tails that depend on $Re$: the low-$F_t$ tails are present at lower $Re$ while high-$F_t$ tails are present at higher $Re$. To our knowledge, this is the first report of $p(F_t)$ in any transitional shear flow. 

We find that the central portions of these PDFs are closely approximated by Fisher-Tippett distributions.
The cumulative distribution function (CDF) of the Fisher-Tippett (also called Generalized Extreme Value) distribution that we will use is:
\begin{equation}
    \label{eq:FisherTippett}
     \mathbb{P}(X\leq h) = P_{FT}(h)\equiv 1 - e^{ - (1 + \xi  (\mu-h)/\sigma)^{-1/\xi}}
\end{equation}
where the location $\mu$, scale $\sigma$, and shape $\xi$ are fitting parameters. Equation \eqref{eq:FisherTippett} is the CDF for minima of a set of random variables, and it is this form that fits our data.
We fit $p(F_t)$ with the Fisher-Tippett density $p_{FT}(h)= d P_{FT} / d h$ shown as dashed curves on figure~\ref{fig:pdf_fit}. 
(The resemblance of the abbreviation FT for Fisher-Tippett and the notation $F_t$ for turbulent fraction is coincidental.)

Figure~\ref{fig:pdf_fit} shows that the central region near the maximum of each PDF fits well with the Fisher-Tippett distribution inside a range spanning from $\hleft$ to $\hright$. 
As an example, these lower and upper bounds of the fit are indicated by colored and white circles for $Re = 830$.
The quality of the fit is particularly good for $Re < 1000$ but shows some noticeable deviations at $Re=1000$ and $Re=1050$.
The fitting parameter values as a function of the Reynolds number are plotted in figure~\ref{fig:FT_param}, which will be discussed below.

The turbulence fraction $F_t$ defined in equation~\eqref{eq:Ft} is not a maximum of a set of independent quantities (although it includes a Heaviside function which, like the maximum, is a non-analytic operation). Hence, it is not obvious that $F_t$ should be governed by an extreme value distribution. 
Even in the case of vorticity maxima, Nemoto \& Alexakis noted that it is not possible to fully justify Fisher-Tippett distributions since vorticity is correlated in space and time and hence the maxima are not independent.
At present we do not have a formal justification for the fits used in figure \ref{fig:pdf_fit} other than that the distributions are clearly non-Gaussian and are fit reasonably well with the Fisher-Tippett form. We hypothesize that the strong spatiotemporal correlations within the localized turbulent bands play a significant role in the statistics, but we leave this for further investigation.
The only way the fits will enter into the analysis that follows is via their parameterisation. In this regard the fits give us a useful representation of the PDFs in terms of three parameters depending on $Re$. It is nevertheless possible that the distributions are of some other type.

The Nemoto \& Alexakis approach requires many numerical simulations of rare events in order to obtain the tails of probability distributions. Here, the AMS approach is particularly useful as it produces large samples of the rare event trajectories that reach destination $\setB$. From the AMS data one can reconstruct the CDF of any observable $X$ depending on a field $\mathbf{u}$ as follows. Each point on a trajectory $\mathbf{u}(t)$ is known to be on a segment from $\setA$ to $\surfS$, from $\surfS$ to $\setA$, or from $\surfS$ to $\setB$. (See figure \ref{fig:Schematic_Markov}.) Hence the CDF can be decomposed into a weighted sum of independent CDFs conditioned on the location of $\mathbf{u}$:
\begin{align}
\label{eq:cdf}
\mathbb{P}(X \leq h ) =\frac{\tau_\mathcal{AS}}{\tau}\,\mathbb{P}(X \leq h ~|~ \mathcal{C}_\mathcal{AS})
+ \frac{\tau_{SA}}{\tau}\,\mathbb{P}(X \leq h ~|~ \mathcal{C}_\mathcal{SA})
+ \frac{\tau_{SB}}{\tau}\,\mathbb{P}(X \leq h ~|~ \mathcal{C}_\mathcal{SB}), 
\end{align}
where $\mathcal{C}_{\mathcal{AS}}$ (resp. $\mathcal{C}_{\mathcal{SA}}$ and $ \mathcal{C}_{\mathcal{SB}}$) is the conditional event that a field $\mathbf{u}$ lies on a trajectory that goes from $\setA$ to $\surfS$ (resp.~from $\surfS$ to $\setA$ or to $\setB$). The weights are the relative time spent in each segment, where
\begin{align*}
\tau & = \tau_{\setA\surfS} + \tau_{\surfS\setA} + \tau_{\surfS\setB} \\
& = \frac{1}{\hat{p}} \,\oT_{\mathcal{AS}}
+ \frac{1-\hat{p}}{\hat{p}} \:\oT_{\mathcal{SA}}
+ \oT_{\mathcal{SB}}.
\end{align*}
We refer the reader back to equation \eqref{eq:tau} for the formula for $\tau$ in terms of $\oT_{\mathcal{AS}}$, etc. The individual CDFs in \eqref{eq:cdf} are obtained in the standard way by rank ordering the sample data and performing a cumulative summation followed by normalisation.

\begin{figure}
    \centering
    \subfloat[]{\includegraphics[width=0.5\textwidth]{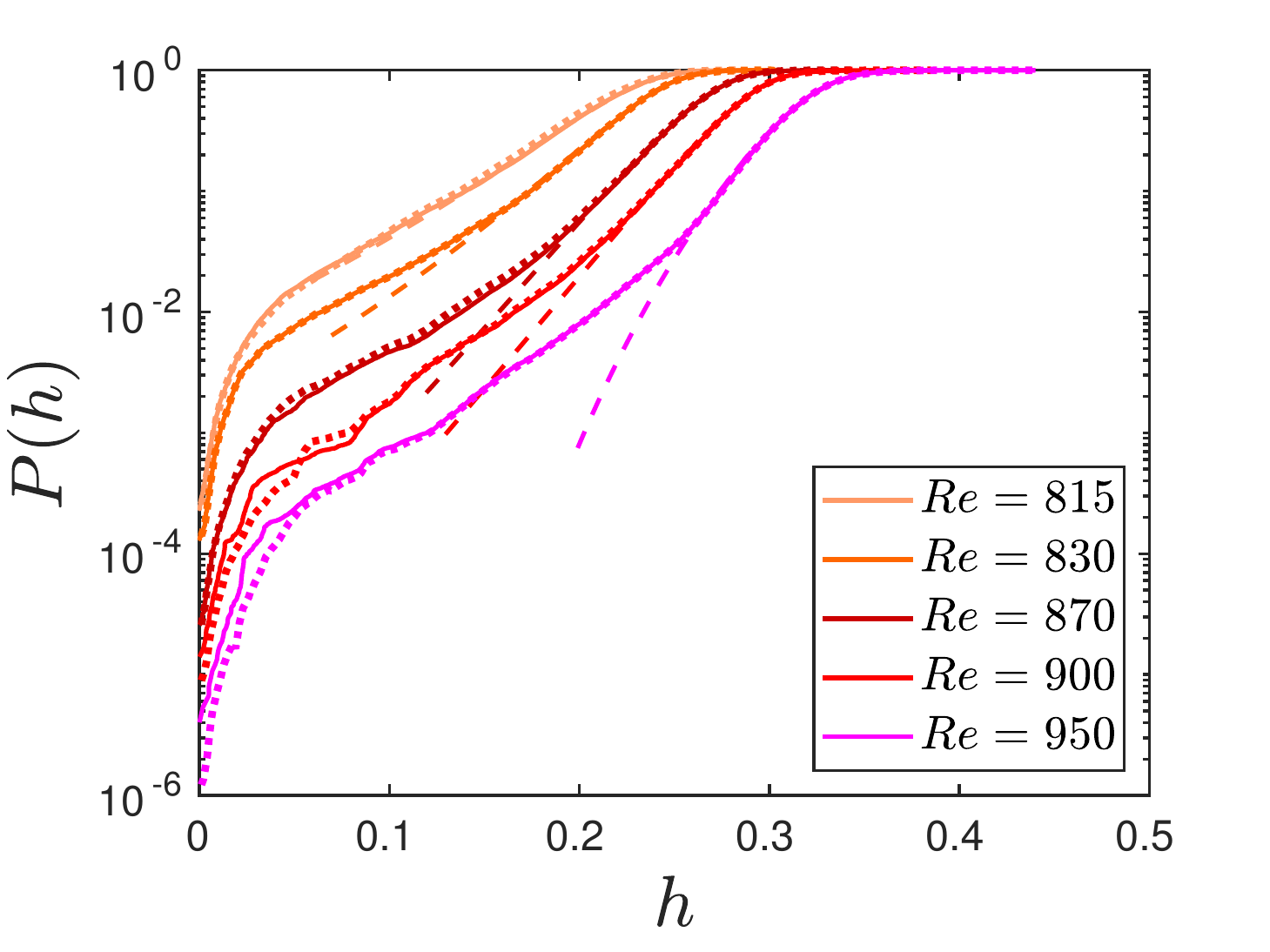}
    \label{fig:cdf_decay}}
    \subfloat[]{\includegraphics[width=0.5\textwidth]{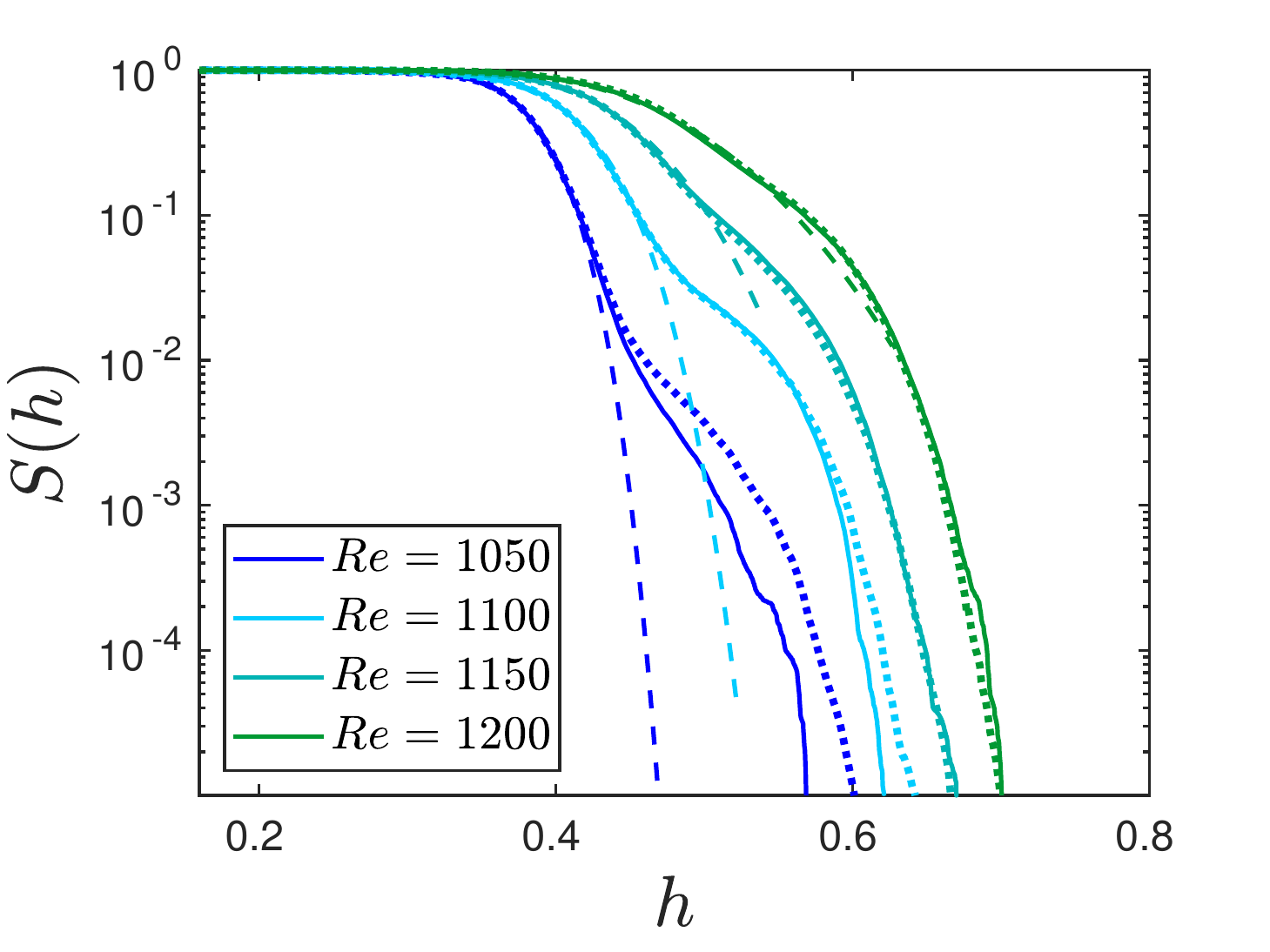}
    \label{fig:cdf_split}}\\
    \subfloat[]{\includegraphics[width=0.5\textwidth]{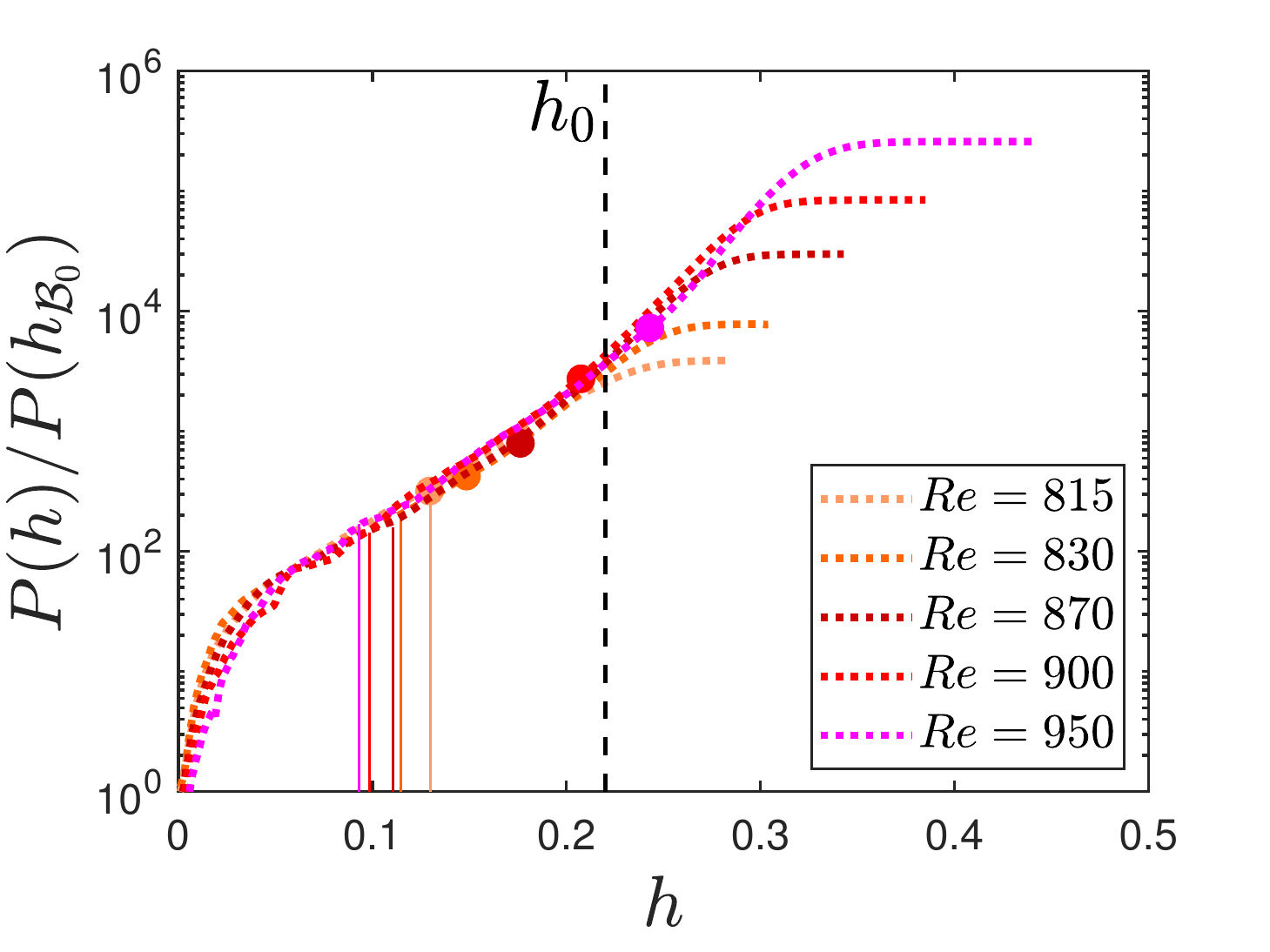}
    \label{fig:cdf_decay_rescaled}} ~
    \subfloat[]{\includegraphics[width=0.5\textwidth]{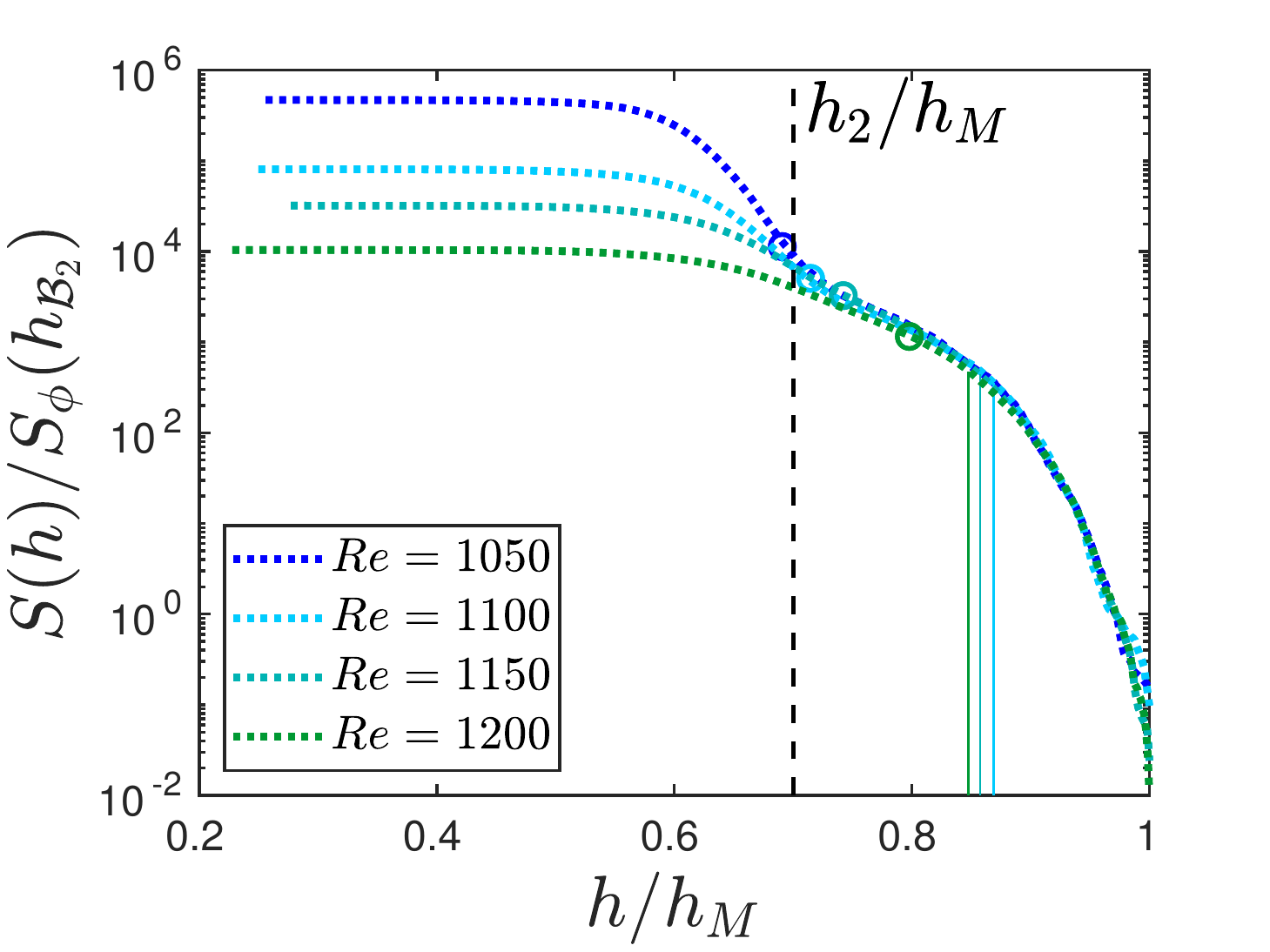}
    \label{fig:cdf_split_rescaled}}
      \caption{(a) Cumulative distribution function $P(h) = \mathbb{P}(F_t \leq h)$ for band decay and (b) survival function $\survival{h} \equiv 1 - P(h) \equiv \mathbb{P}(F_t \geq h)$ for band splitting at values of $Re$ indicated in the legend. Continuous lines are obtained from Monte Carlo and dotted lines are from the AMS algorithm. Dashed lines correspond to fits to a Fisher-Tippett distribution \eqref{eq:FisherTippett}. 
      (c, d) Distributions from the AMS algorithm rescaled by $P(\hBo)$ and $S_\phi(\hBtwo)\equiv 1-P_\phi(\hBtwo)$. In the splitting case (d), the range in $F_t$ is rescaled by $h_M(Re)=\max(F_t)$. Coloured points in (c) show $\hleft$, the lower bounds of the fit to the PDF with a Fisher-Tippett density function (see figure~\ref{fig:pdf_fit}). Similarly the open points in (d) show the upper bounds $\hright$. Vertical lines show the break-even points defined in the text. }
    \label{fig:cdf_Ft}
\end{figure}

Figure \ref{fig:cdf_decay} shows the CDF $P(h) = \mathbb{P}(F_t \leq h)$ for the low-$Re$ decay cases and figure \ref{fig:cdf_split} shows its complement, the survival function $\survival{h} \equiv 1 - P(h) \equiv \mathbb{P}(F_t \geq h)$, for the high-$Re$ splitting cases. 
Results from the Monte Carlo simulations are shown as continuous curves, while those from AMS have been included as dotted curves.
It can be seen that the distribution functions constructed from AMS improve the quality of the tails from Monte Carlo, particularly in the range $900 \le Re \le 1100$ where Monte Carlo systematically underestimates the tails associated with rare transitions. (We note, however, that even with the improvements from the AMS, there remain some sampling effects in the weak tails.) Dashed curves show the Fisher-Tippett CDFs obtained by fitting the PDFs of $F_t$ shown in figure~\ref{fig:pdf_fit}.

\subsection{Timescales from extreme value distributions}

We can now apply the Nemoto \& Alexakis approach~\cite{nemoto2021extreme} to our decay and splitting data. The essential idea is to scale the CDFs and obtain forms that separate into approximately $Re$-independent portions and $Re$-dependent portions that can be fit to Fisher-Tippett distributions. From this it is possible to express the mean timescales for decay and splitting directly in terms of the parameters of the Fisher-Tippett distributions. 

We will first describe the decay case and afterwards summarise the differences for the splitting case. 
Recall that in the decay case the score function for AMS is just the turbulence fraction and the boundary of the laminar state is $\hBo = 0.0001$, meaning that trajectories $\mathbf{u}(t)$ that reach the threshold $F_t(\mathbf{u}) = \phi(\mathbf{u}) = \hBo$ from above are considered to have undergone transition to the laminar state.
As shown in figure~\ref{fig:cdf_decay_rescaled}, by rescaling CDFs by their value at the threshold $P(h_{\setBo})$, the low-probability tails for different $Re$ nearly collapse to a common curve. More specifically, we observe that below a value $\hone$, indicated on the plot, the ratio ${P(h)}/{P(h_{\setBo})}$ depends only weakly on $Re$. (Moreover, some of this dependence is likely due to sampling errors of the low-probability tails.)
Flow fields $\mathbf{u}$ such that $F_t(\mathbf{u})\in[\hBo ,\hone]$, called the \emph{collapse zone} in the following, are in an intermediate state that can either recover (missed decay) or die (successful decay). This process is not a strong function of $Re$. 
Above $\hone$, the rescaled CDFs depend strongly on $Re$, varying by over an order of magnitude over the $Re$ range shown. Significantly, however, for almost all $Re$ this portion of the CDFs lies within the region that is well fit by the Fisher-Tippett distribution. Concretely, the coloured points in figure~\ref{fig:cdf_decay_rescaled} indicate the left-most values of $h$ for each $Re$ for which the Fisher-Tippett fits are good and in almost all cases, these points are below $\hone$, with the point for $Re=950$ slightly above $\hone$.

Following Nemoto \& Alexakis, we can connect the CDFs to decay lifetime $\taud$. The algebraic statement is
\begin{equation}
\taud  
= \frac{\delta t}{P(\hBo)}
= \delta t \frac{P(\hone)}{P(\hBo)} \frac{1}{P(\hone)}
\simeq \underbrace{\delta t \frac{P(\hone)}{P(\hBo)}}_{\Pid}
\underbrace{\frac{1}{1-e^{-(1 + \xi(\mu - \hone)/\sigma )^{-1/\xi}} }}_{f_d(Re)},
\label{eq:superexp_d}
\end{equation}
which we will explain in steps. 

The first equality can be understood as follows~\cite{nemoto2021extreme}. Consider estimating $\tau_d$ by Monte Carlo simulation with $N_{\rm decay}$ independent realisations of decay events. Then $\tau_d = T_{\rm total}/N_{\rm decay}$, where $T_{\rm total}$ is the total combined time to decay for all realisations. Further letting $T_{\rm total} = \delta t \: N_{\rm total}$, where $N_{\rm total}$ is the total number of sample points on all trajectories and $\delta t$ is the sample time, we have $\tau_d = \delta t \: N_{\rm total}/N_{\rm decay}$. Finally, from $N_{\rm decay}$ simulations that terminate at $\hBo$, we have $P(\hBo) = N_{\rm decay}/N_{\rm total}$, since there are $N_{\rm decay}$ out of 
$N_{\rm total}$ sample points with $F_t \le \hBo$. In practice we construct $P(\hBo)$ from AMS simulations via \eqref{eq:cdf} with a sampling time $\delta t = 1$.

The remainder of \eqref{eq:superexp_d} consists of multiplying and dividing by $P(\hone)$ and then applying the previous observations about figure~\ref{fig:cdf_decay_rescaled} to decompose \eqref{eq:superexp_d} into a factor $\Pid$, that depends only weakly on $Re$, and $1/P(\hone)$, that depends strongly on $Re$.
Furthermore, we approximate $P(\hone)$ by the Fisher-Tippett distribution evaluated at $\hone$. The $Re$-dependence of $f_d \simeq 1/P(h_0)$ is contained in the $Re$-dependence of the parameters $\mu$, $\sigma$ and $\xi$. We return to this after discussing the splitting case.

In almost all respects the splitting analysis is the same as that of the decay case. The only important differences comes from the fact that the score function $\phi$ for splitting \eqref{eq:score} is not the turbulence fraction $F_t$. However, $\phi$ and $F_t$ are closely related, both in terms of expression \eqref{eq:score} and in terms of the values they take during band splitting in figure~\ref{fig:score_probe_split}.
A split is deemed to have occurred when $\phi(\mathbf{u}(t))$ reaches $\hBtwo$ from below. Hence, analogously with \eqref{eq:superexp_d}, the time scale for splits is related to the survival function of $\phi$ evaluated at $\hBtwo$: 
\begin{equation}
\taus =  
\frac{\delta t}{\mathbb{P}(\phi > \hBtwo)} = \frac{\delta t}{\survivalphi{\hBtwo}},
\label{eq:tausphi}
\end{equation}
where $S_\phi$ is the survival function for $\phi$. 
While one could analyse distributions of the score function $\phi$, the turbulence fraction is ubiquitous in this field and the distributions in figures~\ref{fig:pdf_fit} and \ref{fig:cdf_split} are of general interest. Hence it is preferable to work with these distributions, even though it will be necessary to rescale the CDF  in figure \ref{fig:cdf_split} using $\survivalphi{\hBtwo}$. 
This is not as awkward as it may seem since $\survivalphi{\hBtwo} = N_{\rm split}/N_{\rm total}$, by the same argument as above for decay. Hence, while we write the normalisation in terms of $S_\phi$, it is not necessary to have access to this CDF to know the normalisation, which is determined simply from the number of sample points and the number of splitting cases. 
To collapse the CDFs we must also rescale the horizontal axis of figure \ref{fig:cdf_split}. We rescale by $h_M$, the maximum value of $F_t$ observed at each $Re$. 
This was unnecessary in the decay case because the minimum value of $F_t$ is achieved at the $Re$-independent termination value $\hBo$.

Figure~\ref{fig:cdf_split_rescaled} shows the rescaled CDFs for band splitting. We observe that the low probability tails for different $Re$ collapse well to a common curve $h \ge h_2$, while for $h < h_2$ the rescaled CDFs depend strongly on $Re$. Also shown as points in figure~\ref{fig:cdf_split_rescaled} are the upper limits for which the curves are well approximated by Fisher-Tippett distributions. These points are above, or nearly above $h_2$ in all cases. Hence, we can again exploit this to approximate the splitting time scale in terms of parameters of the Fisher-Tippett distributions. Starting from \eqref{eq:tausphi} the algebra is
\begin{align} 
\label{eq:superexp_s}
\taus  
= \frac{\delta t}{\survivalphi{\hBtwo}}
= \delta t \,\frac{\survival{h_2}}{\survivalphi{\hBtwo}}
           \frac{1}{\survival{h_2}}
\simeq \:\underbrace{\delta t\, \frac{\survival{h_2}}{\survivalphi{\hBtwo}}}_{\Pis}\:
\underbrace{e^{ (1 + \xi(\mu-h_2)/\sigma )^{-1/\xi} }}_{f_s(Re)}.
\end{align}
We thus obtain an approximation for $\taus$ as a product of a factor $\Pis$, weakly dependent on $Re$, and a factor $f_s(Re)$, strongly dependent on $Re$ via the parameters $\mu$, $\sigma$, $\xi$, as well as $h_2$. Note that $h_2/h_M$ is constant at the start of the collapse zone, but $h_M$ depends on $Re$, and hence so does $h_2$.
Values of $h_2$ and $h_M$, as well as $\hone$, are given in table~\ref{tab:hRe}.

Finally, the vertical lines in figures~\ref{fig:cdf_decay_rescaled} and \ref{fig:cdf_split_rescaled} indicate the break-even point for transition events to take place. 
These have been determined from DNS trajectories that originate in $\setA$ as follows. For a given value of $h$, we compute the fraction of trajectories attaining $F_t = h$ that successfully transition to $\setBo$ or $\setBtwo$, without returning to $\setA$. The value of $h$ for which this fraction is 1/2 is the break-even point. This is conceptually similar to finding where the \emph{committor function} for a stochastic process \cite{vanden2006towards} is equal to 1/2, but here we condition on values of the turbulence fraction and not points in phase space.
At $Re=1050$ we have not obtained a sufficient number of DNS trajectories undergoing transition to $\setBtwo$ to estimate the break-even point, and hence this case is not included in figure~\ref{fig:cdf_split_rescaled}. We provide context for these break-even points in the next section. 


\subsection{Super-exponential scaling}

\begin{figure}[h!]
    \centering
    \subfloat[]{\includegraphics[width=0.5\textwidth]{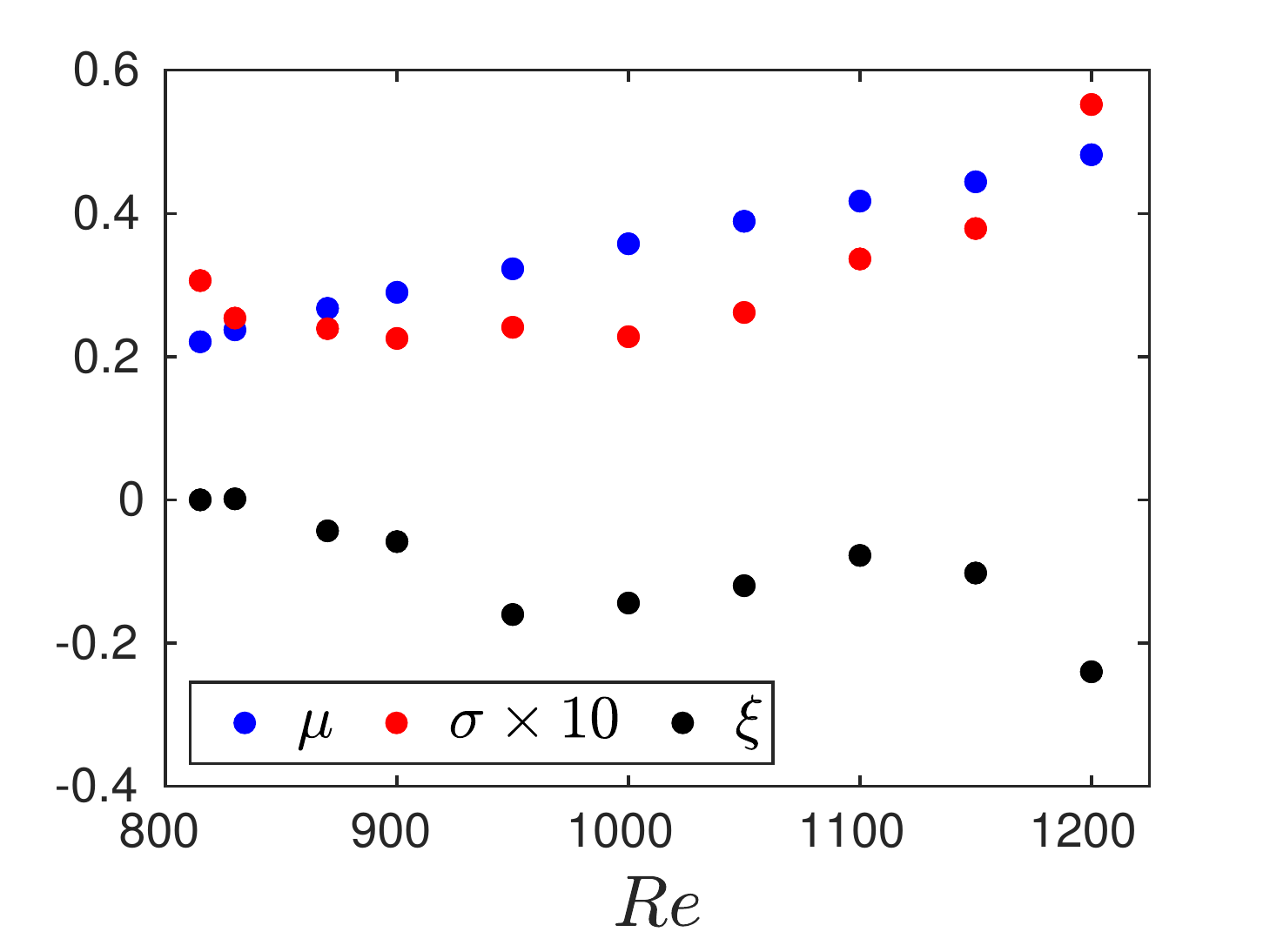}
    \label{fig:FT_param}}
    \subfloat[]{\includegraphics[width=0.5\textwidth]{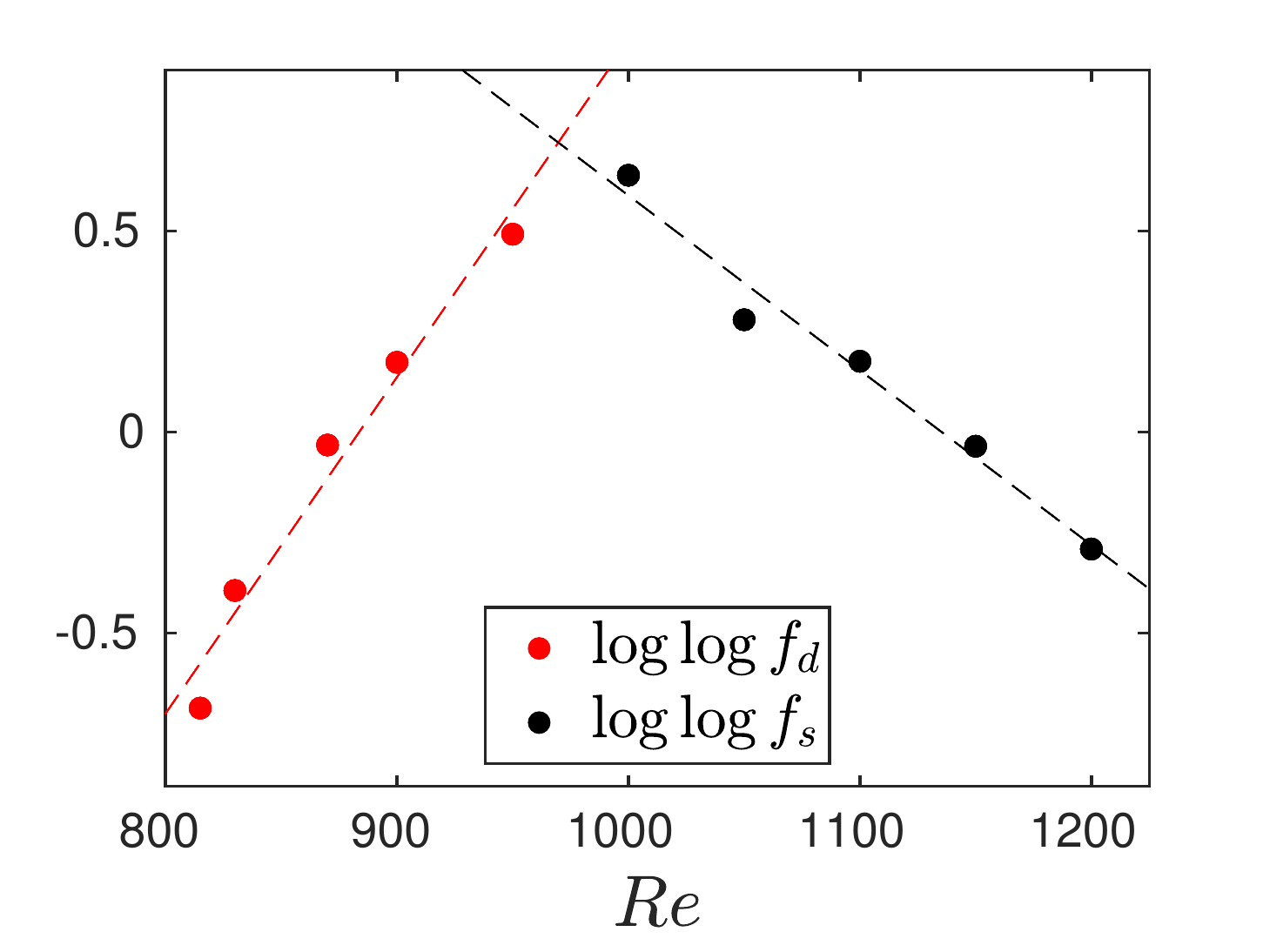}
    \label{fig:FT_loglog}}
\caption{(a) Dependence of the three Fisher-Tippett parameters on $Re$. These have been obtained by fitting Fisher-Tippett distributions to the numerical PDFs $p(F_t)$ over ranges $\hleft \le F_t \le \hright$ as seen in figure~\ref{fig:pdf_fit}.
(b) Dependence of $\log \log f_d$ \eqref{eq:superexp_d} and $\log \log f_s$ \eqref{eq:superexp_s} on $Re$ using the parameters from (a). Dashed lines show linear fits.}
\end{figure}

We now explore the connection between the observed super-exponential dependence of mean lifetimes on $Re$ seen in figure~\ref{fig:tau} and the approximations to the mean lifetime given in \eqref{eq:superexp_d} and \eqref{eq:superexp_s}.
We have argued that the dominant dependence of mean lifetimes on $Re$ is captured by the dependence of the functions $f_d$ and $f_s$ on $Re$. These functions depend on $Re$ via the Fisher-Tippett parameters $\mu$, $\sigma$, and $\xi$ of \eqref{eq:FisherTippett} which are shown in figure~\ref{fig:FT_param}. 
The location parameter $\mu$ varies linearly with $Re$, a feature which can already be seen in the $Re$-dependence of the maxima in figure \ref{fig:pdf_fit}. The $Re$-dependence of the scale $\sigma$ and the shape $\xi$ is less clear; their fluctuations may be due to their sensitivity to the fitting procedure. Since the quality of the fits in figure \ref{fig:pdf_fit} is not improved by the inclusion of more simulation data, the fluctuations may indicate that $p(F_t)$ is not exactly of Fisher-Tippett form even near its maximum.

The parameter $\xi$ plays an essential role in the family of Fisher-Tippett distributions, dividing them into three categories. Those with $\xi>0$ are the Fr\'echet distributions (also known as type II extreme value distributions), while $\xi<0$ corresponds to Weibull (type III). Figure \ref{fig:FT_param} shows that the central portions of most of the curves in figure \ref{fig:pdf_fit} are best fit to Weibull distributions ($\xi$ may be positive for $Re=815$ and 830, but there is too much uncertainty in our fits to be sure).
The limiting case $\xi=0$ is the family of Gumbel distributions (type I), which will play a role in section \ref{sec:pathways}.

Figure \ref{fig:FT_loglog} shows $\log \log f_d$ and $\log \log f_s$ from expressions \eqref{eq:superexp_d} and  \eqref{eq:superexp_s} as a function of $Re$ using the numerically obtained parameter values for each $Re$. Linear fits show that $\log \log f_d \simeq a_d \, Re + b_d$ and $\log \log f_s \simeq a_s \, Re + b_s$ over a range of nearly 200 in $Re$ in each case. Hence both $f_d$ and $f_s$ depend super-exponentially on $Re$ and are at least approximately of the form $[\exp(\exp (a ~ Re + b))]$. 
Given the functional forms of $f_d$ and $f_s$ and the complicated dependence of the fitting parameters on $Re$, the double exponential dependence on $Re$ is only an approximation. Nevertheless, we clearly observe a faster than exponential dependence on $Re$ resulting from modest variation with $Re$ of parameters of the Fisher-Tippett distribution characterising the fluctuations in the one-band state. 

The interpretation of these results comes from the mechanism proposed by Goldenfeld et al. \cite{goldenfeld2010extreme} and subsequently refined by Nemoto and Alexakis \cite{nemoto2018method,nemoto2021extreme}. We focus on the decay case, but similar statements apply to the splitting case. 
The picture is that the statistics of strong turbulent fluctuations are governed by extreme value distributions and this gives rise to the strong $Re$ dependence of the probability $P(\hone)$ of states being in the collapse zone $h \le \hone$; see figure \ref{fig:cdf_decay_rescaled}. Note that most trajectories that enter the collapse zone do not decay directly, but instead return to the one-band state $\setA$. Only when trajectories achieve values of $F_t$ below the break-even points (shown as vertical lines in the figure) are trajectories more likely to decay than to return to $\setA$. 
The probability of decay becomes one at $\hBo$, since this defines the boundary we have chosen for the laminar state $\setBo$, and the rate of ultimate decay is given by $P(\hBo)$ which is much less than $P(\hone)$.
However, the ratio $P(\hone)/P(\hBo)$ is nearly independent of Reynolds number. Hence up to a $Re$-independent multiplicative factor, the decay rate is determined from probability $P(\hone)$. The reason why the CDFs for different $Re$ collapse over a range of turbulence fractions, and why this occurs for both decay and splitting processes, remains unexplained.

We end this section with a few observations and caveats.
We observe that PDFs of $F_t$ are well fit near their maxima by Weibull distributions,  at least for most of the $Re$ range investigated. This is distinctly different from the Fr\'echet distributions observed by Nemoto \& Alexakis for maximum vorticity in pipe flow \cite{nemoto2021extreme}. We note also that while $F_t$ is a non-smooth function of the flow field, it is not given as an extremum over any feature of the field.

The purpose of decomposing the mean lifetimes \eqref{eq:superexp_d},~\ref{eq:superexp_s}) and using the Fisher-Tippett parameter fits is not to obtain quantitatively accurate formulas for $\taud$ and $\taus$, but to gain insight into the source of the super-exponential dependence on $Re$. In this regard we note that the biggest issue, both quantitative and conceptual, with this approach is that we rely on the existence of delimiters $\hone$ and $h_2$ that are simultaneously within the collapse zones and within the range in which the distributions are close to Fisher-Tippett form. As can be seen in figures~\ref{fig:cdf_decay_rescaled} and \ref{fig:cdf_split_rescaled}, this does not hold for $950 \lesssim Re \lesssim 1050$. This was also observed for puff decay in pipe flow: figure 10(a) in \cite{nemoto2021extreme}. This does not invalidate the connection between extreme value statistics and the super-exponential scaling, but it does mean that there is a gap in using the Fisher-Tippett approximation at large time scales that at present we do not see how to close.


\section{Transition pathways}
\label{sec:pathways}

Extreme value theory not only relates the super-exponential scaling of mean lifetimes to the distribution of fluctuations of the one-band state, it also provides a framework for understanding the rare pathways from one state to another. 
In a previous publication~\cite{gome2020statistical} we observed that the dynamics of band splitting were concentrated around a most-probable pathway in the phase space of large-scale Fourier coefficients.
This motivates us to explore connections with \emph{instantons} in the framework of Large Deviation Theory for systems driven by weak random perturbations. See for example \cite{touchette2011basic,grafke2015instanton,grafke2019numerical} and references therein.
The concept is easily illustrated with the following stochastic differential equation
\begin{equation}
    \dot X = -\nabla V(X) + \sqrt{\varepsilon} \eta,
    \label{eq:SDE}
\end{equation}
where $X \in \mathbb{R}^d$, $V$ is a potential, $\varepsilon$ is a perturbation strength and $\eta$ is Gaussian white noise. 
We assume that $V$ has two local minima $\setA$ and $\setB$ separated by a saddle point and we consider transitions from $\setA$ to $\setB$.
In the weak-noise limit $\varepsilon \to 0$, transitions will be rare and the trajectories associated with these rare events will be concentrated around a most probable path that connects states $\setA$ and $\setB$. This is the \emph{instanton}. The dynamics along the instanton is such as to climb uphill from $\setA$ to the saddle point under the influence of weak noise, and then to fall deterministically from the saddle to $\setB$. 

Examples of instantons in fluid systems are found for shocks in Burgers equations \cite{grafke2013instanton, grafke2015instanton}, and have been predicted and experimentally observed in rogue waves \cite{dematteis2019experimental}. The concentration of transition paths around an instanton in a high-dimensional fully turbulent system was first observed by Bouchet \emph{et al.} \cite{bouchet2019rare} in a 2D barotropic model of atmospheric dynamics. Schorlepp \emph{et al.} \cite{schorlepp2021spontaneous} have used instanton calculus to investigate the most likely configurations to generate large vorticity or strain within turbulence in the 3D Navier-Stokes equations.
This phenomenology can also apply to deterministic chaos, as in the solar system \cite{woillez2020instantons}. 
Rolland has discussed instantons specifically in relation to turbulent-laminar transition, both in a model system~\cite{rolland2018extremely} and in plane Couette flow~\cite{rolland_2022}. 

\begin{figure}[h!]
    \centering
    \includegraphics[width=1\textwidth]{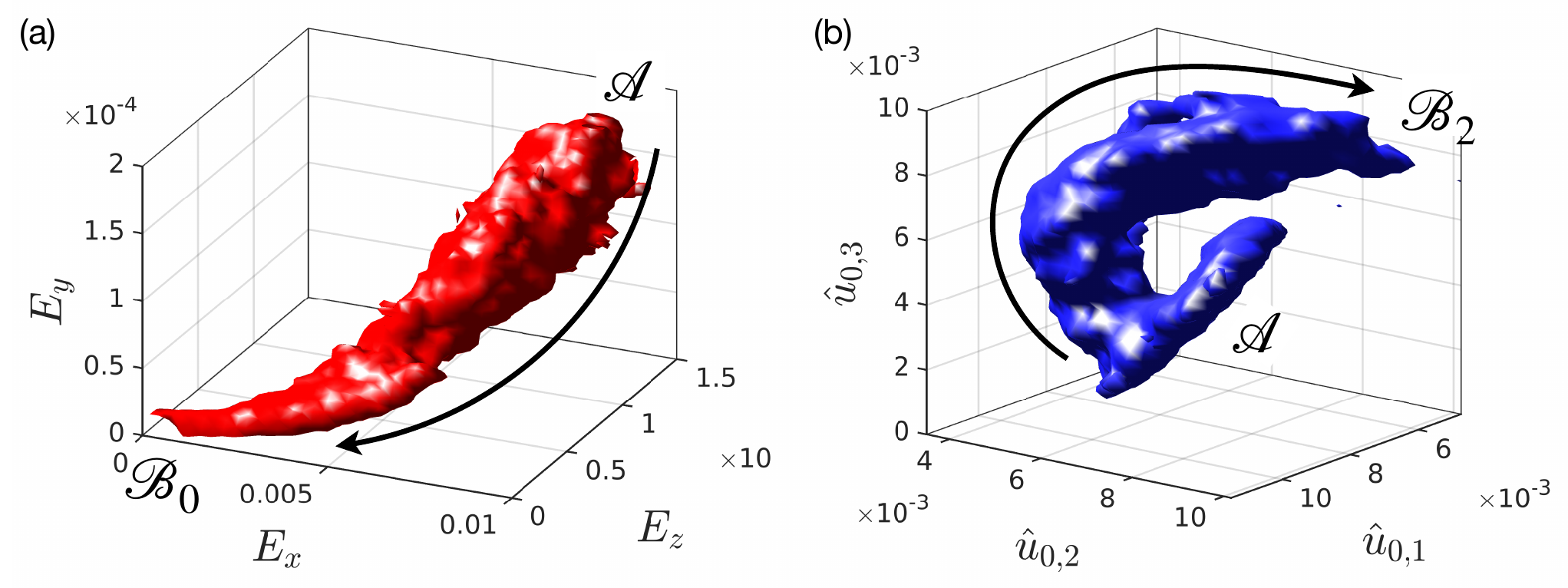}
    \caption{Joint probability density functions for reactive trajectories going from (a) $\setAone$ to $\setBo $ at $Re = 830$ (decay events) or (b) from $ \setAone$ to $\setBtwo$ at Re = 1150 (splitting events). 
    (a) Isosurface of $p(E_x, E_y,E_z)$ enclosing 90\% of the total probability. (b) Isosurface of $p(\hat{u}_{0,1},\hat{u}_{0,2},\hat{u}_{0,3})$ enclosing 80\% of the total probability. 200 trajectories are computed in each case. 
    }
    \label{fig:instanton}
\end{figure}

Rare transitions of the type considered here could exhibit instanton-type behaviour if turbulent fluctuations were to play the role of weak noise. A detailed investigation is outside the scope of this paper, but the current interest in the topic and the capacity of AMS to generate large numbers of rare transitions motivates us to briefly present transition paths for decays and splits. 
Examples of each are shown in figure~\ref{fig:instanton}. By binning samples from 200 transition paths we construct PDFs and then render isosurfaces of these PDFs to reveal the reactive tubes where paths concentrate. 
We include only reactive trajectories that leave $\setA$ and terminate at
the boundary of $\setBo$ or $\setBtwo$ without returning to $\setA$.

The coordinates used for the PDF are chosen separately for decay and splitting. For decay, we show the decay of energy associated with the three velocity components of the flow, $E_x,E_y,E_z$ 
$$
E_{(x,y,z)} \equiv \frac{1}{L_x L_y L_z}\int_{\domain} \frac{1}{2}(u^2,v^2,w^2) \mathrm{d}\mathbf{x}.
$$
Figure~\ref{fig:instanton}(a) shows that the reactive pathway from $\setA$ to $\setBo$ is such that $E_y$ decays most quickly, followed by $E_z$, followed by $E_x$ so that the tube of reactive trajectories approaches $\setBo$ almost tangent to the $E_x$ axis. This ordering of decay of energy components has been reported previously \cite{gome2020statistical,liu2021decay}; here the 90\% probability isosurface shows that almost every successful decay trajectory follows a similar path.   

For splits, we use coordinates similar to those in \cite{gome2020statistical}, the first three $z$ Fourier components $\hat{u}_{0,1},\hat{u}_{0,2},\hat{u}_{0,3}$ of $u$, averaged in $x$ and $y$:
$$
    \hat{u}_{0,n} = \frac{1}{L_x L_z}\int dy \left| \:\int dx \int dz\; u(x,y,z) e^{-2\pi i n z/L_z} \:\right|.
$$
Figure~\ref{fig:instanton}(b) shows that the reactive pathway from $\setA$ to $\setBtwo$ for the case of splits consists of a highly curved tube. This shape arises from the non-monotonicity of the splitting trajectories in these coordinates, as seen in \cite{gome2020statistical}. While a one-band state in $\setA$ is characterized by high $\hat{u}_{0,1}$, the magnitude of $\hat{u}_{0,2}$ decreases at the beginning of a split before reaching its ultimate higher value in the two-band state in $\setBtwo$.

\begin{figure}[h!]
    \centering
    \subfloat[]{\includegraphics[width=0.5\textwidth]{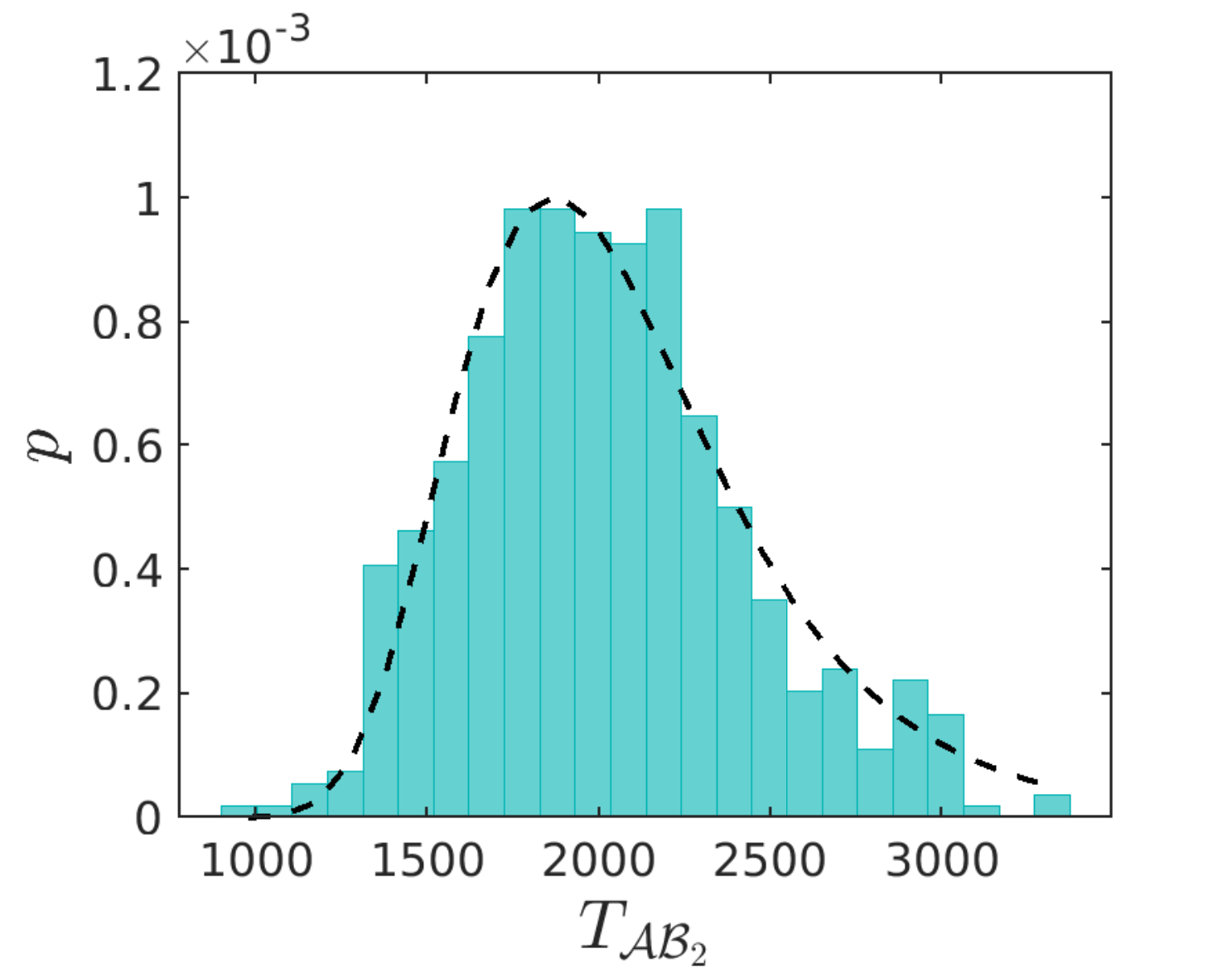}
    \label{fig:pdf_Tab}} ~
    \subfloat[]{\includegraphics[width=0.5\textwidth]{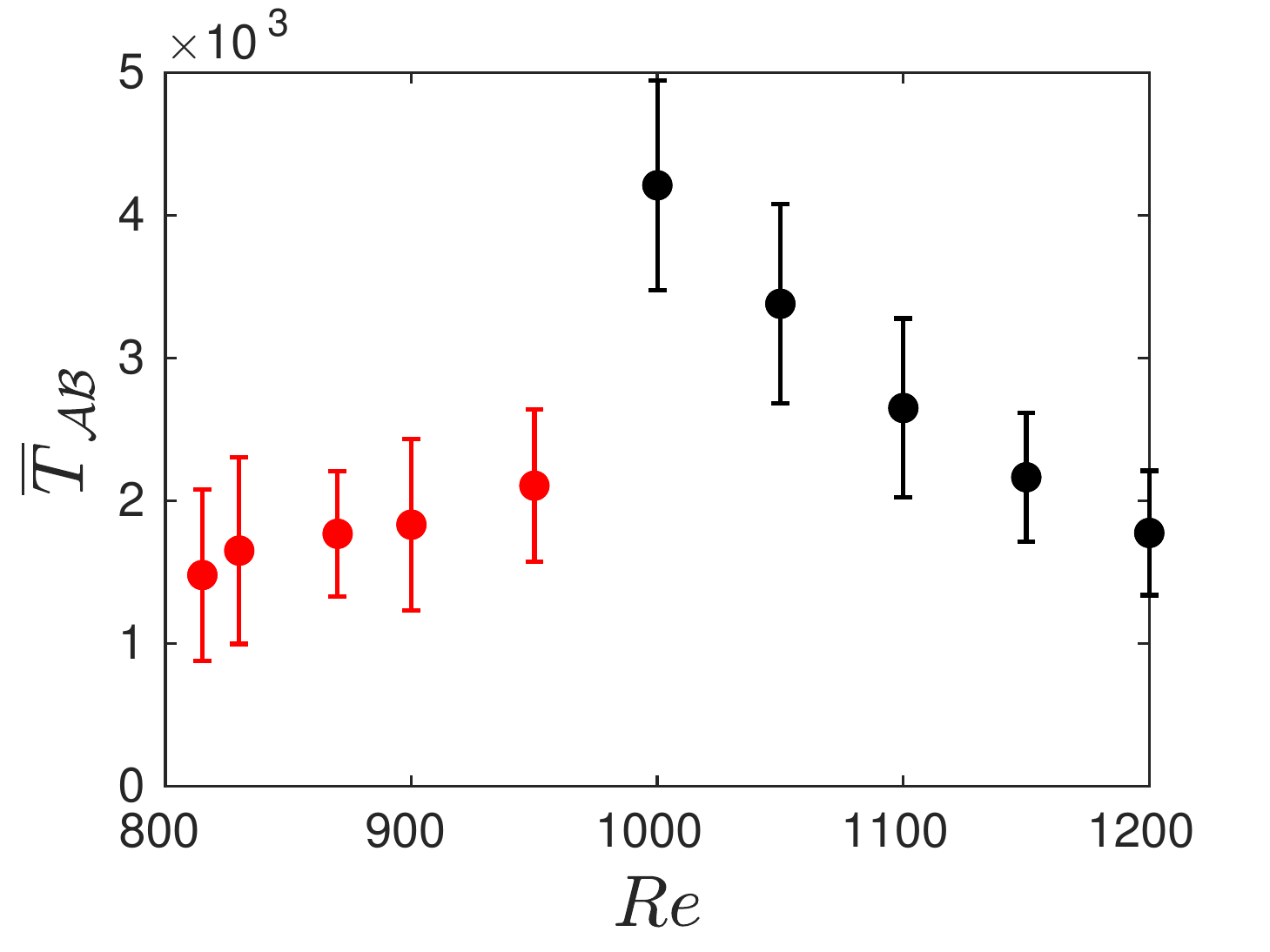}
    \label{fig:Tab}} 
    \caption{(a) Histogram of the reactive times $T_{\setAone \setBtwo}$ at $Re=1150$, estimated with the AMS on $N=500$ trajectories. Dashed lines show a fit with a Gumbel distribution \eqref{eq:Gumbel}
    with $\alpha=1.9\times10^{3}$ and $\beta = 2.7\times10^{-3}$.
    (b) Mean reactive times $\overline{T}_{\setAone \setBo}$ and $\overline{T}_{\setAone \setBtwo}$, for different $Re$, estimated with the AMS. Error bars indicated one standard deviation. Reactive times are measured from a random point in $\setAone$ to the boundary of $\setBo$ or $\setBtwo$. 
    }
    \label{fig:reactive_time}
\end{figure}

The transition pathways can also be described by the distribution of reactive times $T_{\mathcal{AB}}$.
Reactive times have been characterised by Gumbel distributions
\begin{equation}
    p_{\rm Gum}(T) = \beta e^{ -\beta (T - \alpha)} \exp{(- e^{-\beta (T - \alpha)} )},
    \label{eq:Gumbel}
\end{equation}
rigorously in simple stochastic ODEs in the weak noise limit \cite{cerou2013length}, and observationally in one-dimensional stochastic PDEs \cite{rolland2016computing,rolland2018extremely} and in the decay of uniform turbulence in the Navier-Stokes equations \cite{rolland_2022}. 
We find that the distributions of reactive times $T_{\setAone \setBo}$ for decays and $T_{\setAone \setBtwo}$ for splits are consistent with Gumbel distributions for each $Re$ and hence also with instanton-like behaviour.
Figure~\ref{fig:pdf_Tab} illustrates this for $Re=1150$, but the relatively small number of computed reaction trajectories (around 500 for this $Re$) precludes drawing more definite conclusions. 
The mean duration of reactive trajectories and their standard deviation as a function of $Re$ are shown in figure~\ref{fig:Tab}. The mean reactive times $\overline{T}_{\setA\setB}$ vary only modestly with $Re$ within each of the decay and the splitting regimes, as do the standard deviations (shown by the error bars). 

The results presented in this section were motivated by interest in rare-event pathways and instantons in particular. We observe that reactive trajectories for both decays and splits concentrate around a reactive tube in phase space. This suggests that turbulent fluctuations are dominated by the collective behaviour of trajectories along a most-probable path, which may be an instanton.
We observe mild contraction of pathways as we vary $Re$ and events become rare.
(See Supplementary Material.) Such contraction would be expected if the transitions exhibited instanton-like behavior. At the present time, even using the AMS algorithm, we have not produced sufficient numbers of independent reactive trajectories at very high transition times to draw definite conclusions and more work is needed to relate this behaviour to the Large Deviation theory.

\section{Discussion}

Determining -- or even defining -- the threshold for turbulence in wall-bounded shear flows has been an important question since Reynolds' 1883 article \cite{Reynolds:1883}. Over time it has become clear that transitional turbulence is typically metastable and that transitions from metastable states play a crucial role in determining the onset of sustained turbulence \cite{pomeau,bottin1998statistical,bottin1998discontinuous,peixinho2006decay,hof2006finite,willis2007critical}. 
The culmination of this realization was the study of Avila {\em et al.}~\cite{avila2011onset} that determined the mean lifetimes for puff decay and puff splitting in pipe flow and showed that these lifetimes cross at a critical value of the Reynolds number $Re_c$. Although this work involved both numerical simulations and experiments, it was only through experiments that the very long lifetimes associated with $Re_c$ were accessible. 
This has driven interest in capturing transitions from long-lived metastable states in wall-bounded flows via numerical simulations in order to obtain a clearer theoretical understanding of these events and of their Reynolds number dependence. 

We have used the Adaptive Multilevel Splitting algorithm \cite{cerou2007adaptive,cerou2011multiple,cerou2019adaptive} to obtain rare events in plane channel flow. We have specifically analysed transitions from the metastable one-band state to either laminar flow (decay) or to a two-band state (splitting) in tilted-domain simulations of the 3D Navier-Stokes equations with $2 \times 10^7$ degrees of freedom. Using AMS on this large system we have been able to obtain mean lifetimes as large as $5 \times 10^6$ in advective time units with a gain in computational efficiency of a factor of up to 100 over the standard Monte Carlo approach. This has permitted us to access timescales near the lifetime crossing point for this flow. With the significant number of rare transitions we obtained, we have been able to construct weak tails in the probability distribution functions for the turbulence fraction. Exploiting ideas by Goldenfeld, Gutenberg \& Gioia \cite{goldenfeld2010extreme} and Nemoto \& Alexakis \cite{nemoto2018method,nemoto2021extreme}, we have been able to link directly the super-exponential variation of mean lifetimes with $Re$, for both decay and splitting, to the distribution of fluctuations in the one-band state. Finally, we have examined briefly the reaction pathways for decay and splitting. 

Without conducting an extensive companion study in a large untilted domain, we cannot rule out effects of our narrow tilted domain on the transition rates and paths. However, we can cite comparisons of thresholds in the two types of domains. Shimizu~\& Manneville \cite{ShimizuPRF2019} carried out channel flow simulations in large domains of size $L_{x^\prime}\times L_{z^\prime}=500 \times 250$ or $1000 \times 500$ and obtained a threshold between $Re=905$ and 984 for one of the two regimes they studied. This is quite close to the crossover at $Re\approx 980$ between the decay and splitting times that we have computed here in a narrow tilted domain via AMS. In plane Couette flow, the threshold for transition to turbulence was estimated to be $Re=325$ by Shi \emph{et al.}~\cite{shi} as the decay-splitting lifetime crossing in computations in a narrow tilted domain. This is the same as the value estimated experimentally by Bottin \emph{et al.}~\cite{bottin1998statistical,bottin1998discontinuous} and numerically by Duguet \emph{et al.}~\cite{duguet2010formation}, in rectangular domains of size $380 \times 70$ and $800 \times 356$.  An experiment in a much larger domain of size $3927 \times 1500$ by Klotz et al.~\cite{klotz2022phase} yields $Re=330\pm 0.5$ as the threshold .

Throughout this study we have focused on the turbulence fraction as a scalar observable of the state of the system, in large part because it is an easily obtainable quantity of general interest. While turbulence fraction is presumably not a mechanistic driver of either event, it is a very informative observable that is highly correlated to the distance to the targeted states. 
Our analysis of the super-exponential dependence of mean lifetimes on $Re$ is probabilistic and relies heavily on the observed, but unexplained, collapse of rescaled distributions of $F_t$ over what we call the collapse zone.

This approach is complementary to the dynamical-systems approach to turbulence \cite{eckhardt2007turbulence,kerswell2005recent,kawahara2012significance,graham2021exact}.
It would be useful to connect these approaches and to understand the mechanisms at work within the collapse zone. A particular question is the role played by saddle points or edge states \cite{eckhardt2007turbulence,schneider2007turbulence,duguet2008transition,chantry2014studying,paranjape2020oblique} in creating behaviour that can be rescaled and collapse to $Re$-independent form, because this is a key ingredient in how turbulent fluctuations are connected to decay and splitting events. 
While there is much previous work on decay from a dynamical-systems perspective, there is little to rely upon in the case of splitting. 

Our investigation of reaction pathways demonstrates their concentration in phase space for both decay and splitting events. We have also observed a Gumbel distribution for the reaction times. The mild contraction of pathways that we have observed as the transition probability becomes very low resembles an instanton, but is inconclusive.
To better support this picture, we would need to quantify the level of the fluctuations of the effective degrees of freedom in the system and how the fluctuation levels depend on the Reynolds number. Following this, we would need to compare the transition-rate dependence on the Reynolds number to what would be expected from the level of fluctuations within Large Deviation theory.
This would require us to disentangle the effect of $Re$ on turbulent fluctuations from its effect on the  potential term, which itself strongly depends on Reynolds number as seen by the parameterisation of the PDFs within the one-band state (figures \ref{fig:pdf_fit} and \ref{fig:FT_param}). This approach would thus require the computation of the action minimizer in Large Deviation theory, which is out of the scope of the current study.
This fundamental issue is related to the absence of a second parameter that would independently control the level of turbulent fluctuations and thereby allow an approach to a low-noise limit.
We note that the states studied here are localised and insensitive to domain length. Hence domain size, the one parameter other than $Re$ available in the numerical simulations, does not provide a means to influence the effect of fluctuations on the transitions. 
We refer the reader to the important studies of Rolland \cite{rolland2018extremely,rolland_2022} on rare events in transitional shear flows. 

Finally, while we have succeeded in using the AMS algorithm to compute rare events in the 3D Navier-Stokes equations represented by $O(10^7)$ degrees of freedom, the experience has not been without difficulties. The most notable issues are: (1) the algorithm sometimes stagnates, making very slow progress toward obtaining trajectories reaching the target state and (2) the variance in the estimated mean lifetimes associated with the AMS realisations is large, thus requiring the costly step of running multiple realisations. The method used here could possibly be improved with the implementation of Anticipated AMS \cite{rolland_2022}. 
Most importantly, the score function is well known to be crucial to efficient performance of the algorithm. Finding a score function that targets successful splitting events has been particularly challenging. Although we have obtained a serviceable empirical score function based largely upon the turbulence fraction, a more far-ranging search for appropriate score functions is needed.

\enlargethispage{20pt}

\dataccess{The data that support the findings of this study were generated using the open source code Channelflow \cite{channeflow} and are available from the authors upon request.}

\aucontribute{SG conceived of and carried out all simulations and data analysis. SG, LST and DB interpreted the results and wrote the paper.}

\competing{Authors declare no competing interests}

\funding{This work was supported by a grant from the Simons Foundation (Grant number 662985, NG).}

\ack{The calculations for this work were performed using high performance computing resources provided by the Grand Equipement National de Calcul Intensif at the Institut du D\'eveloppement et des Ressources en Informatique Scientifique (IDRIS, CNRS) through grants A0082A01119 and A0102A01119.  We wish to acknowledge Anna Frishman, Tobias Grafke, Freddy Bouchet, Takihiro Nemoto and Alexandros Alexakis for helpful discussions. We also thank Florian Reetz and Alessia Ferraro for their assistance in using Channelflow. 
This work is dedicated to the memory of our dear friend and colleague Charlie Doering.}

\newpage
\begin{center}
  {\bf \large Appendices}
  \end{center}

\appendix

\section{Impact of perturbation level and sample size on the variance of the AMS}

Estimating rare events with the AMS (Adaptive Multilevel Splitting) algorithm for a high-dimensional system such as ours is a trade-off between accuracy of the estimate and computational cost. 
It is known from previous studies on low-dimensional systems \cite{rolland2015statistical, rolland2018extremely} that the variance of the AMS scales with sample size $N$ like $1/\sqrt{N}$, and that completely unbiased results depend, among other things, on the definition of the score function and the number of degrees of freedom. In \cite{rolland2018extremely} it was shown empirically that
$|\pAMS - \pMC |/ \pMC$ scales as $1/N$, where $\pAMS$ and $\pMC$ are the transition probabilities estimated by the AMS and MC (Monte Carlo) methods, respectively

Although a large sample size $N$ is desirable to produces low variance, sample sizes larger than $N=100$ are challenging in terms of computational time and memory in our case.
If smaller sample sizes are used, the accuracy of the estimator can be improved using multiple AMS realisations. 
We have verified the evolution of the AMS estimator $\hat{p}$ for different values of $N$ and $\epsilon$ in Table~\ref{tab:ams_N_eps} and find that good agreement with $\pMC$ was achieved at $N=50$. We thus decide to take $N=50$ for all $Re$, and we further average results over $N_{\text{AMS}}$ realisations as listed in the main paper. 

\begin{table}[h!]
    \centering
    \begin{tabular}{|c|c|c|c|c|}
        \hline
          $\epsilon$ & $N$  & $\pAMS$ & $|\pAMS-\pMC|/\pMC$ & $\sigma(\pAMS)$ \\
         \hline
         $1\times10^{-5}$ & 30 & 0.0351 & 0.250 &	 0.0365 \\
         \hline
          $1\times10^{-5}$ & 50 & 0.0456 & 0.021 &  0.0197 \\
         \hline
          $1\times10^{-5}$ & 100 & 0.0451 & 0.032 &  0.0137 \\
         \hline
         $5\times10^{-6}$ & 50 & 0.0455 & 0.022 &	 0.0299 \\
         \hline
         $1\times10^{-6}$ & 50 & 0.0487 & 0.047 &	 0.0085 \\
         \hline
    \end{tabular}
    \caption{Dependence of the accuracy of the AMS estimator of the probability of transition $\pAMS$ on $N$ and on $\epsilon$, at $Re=1150$, where $\pMC = 0.047$ has been obtained from Monte Carlo simulations. $\sigma(\pAMS)$ is the standard deviation of $\pAMS$ from  at least $N_\text{AMS} = 5$ AMS realisations.}
    \label{tab:ams_N_eps}
    \vspace*{-4pt}
\end{table}

We verify that the estimators of the transition probability $\hat{p}$ are not strongly dependent on $\epsilon$ (see Table~\ref{tab:ams_N_eps}) for $Re=1150$. We observe that if the perturbation strength is too low ($\epsilon=10^{-6}$), the deviation from $\pMC$ slightly increases, because of a lesser diversity of trajectories.
The perturbation must leave the score function unchanged at the cloning time, otherwise it can alter the trajectory selection process. It is also possible that the perturbation alters the trajectories and their statistics, compared to a fully deterministic strategy such as Monte Carlo.
The effect of perturbation level $\epsilon$ used in cloning trajectories is an issue when the transition probability is very low. Low perturbation levels can lead to low diversity of the clone samples, and thus a stagnation of the iterative process. On the other hand, the perturbation at the cloning time must be high enough so that the average time to return to $\setAone$ or to reach $\setBtwo$ is larger than the Lyapunov time of the system. 
For each $Re$, $\epsilon$ is then chosen as the minimal stochastic input that promotes trajectory diversification and for which the algorithm does not stagnate.

\section{Evolution of reactive tubes with the Reynolds number}

\begin{figure}
    \centering
    \subfloat[]{\includegraphics[width=0.5\textwidth]{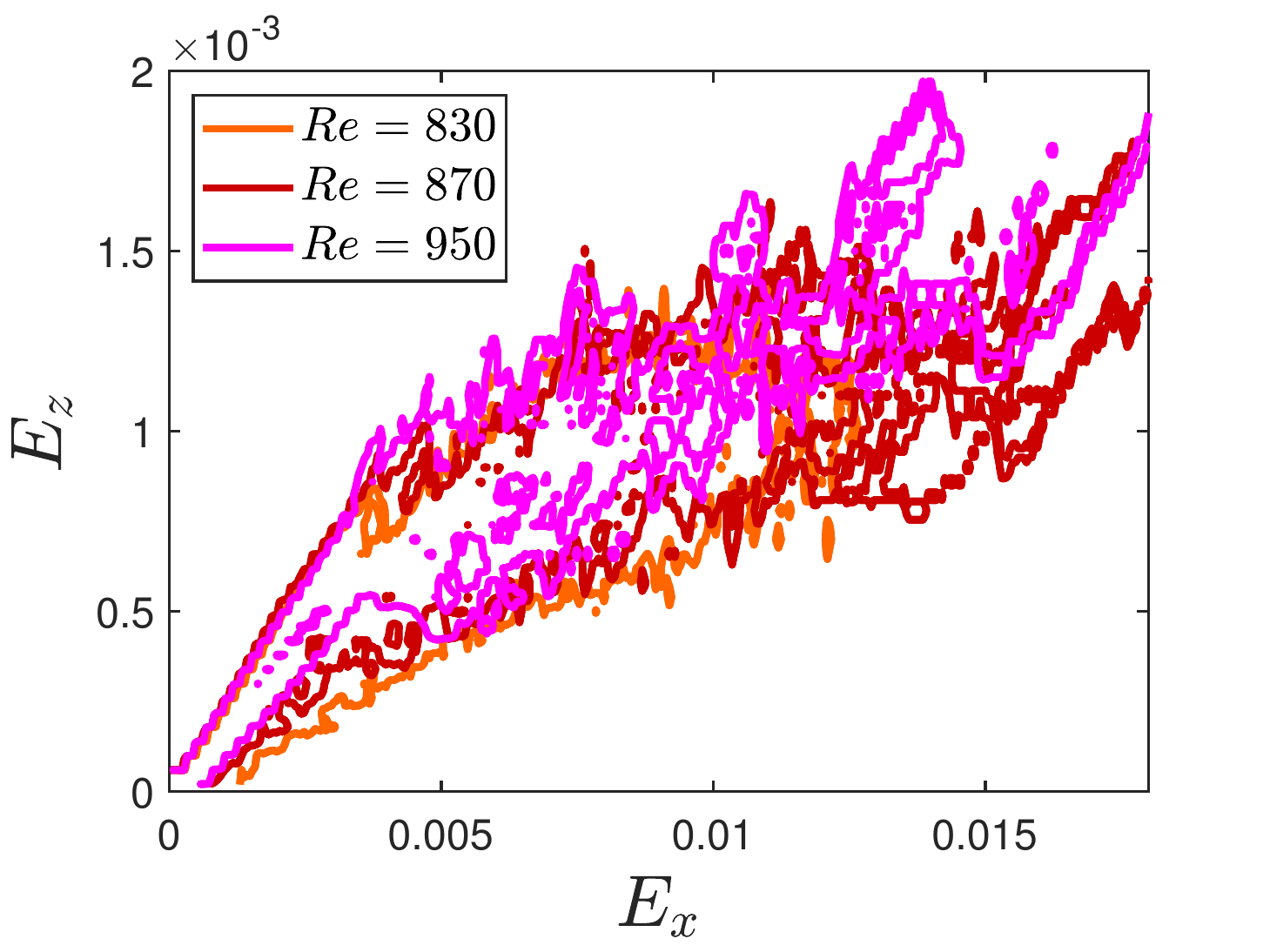}
    \label{fig:instanton_pdf_Re_decay}}
      \subfloat[]{\includegraphics[width=0.5\textwidth]{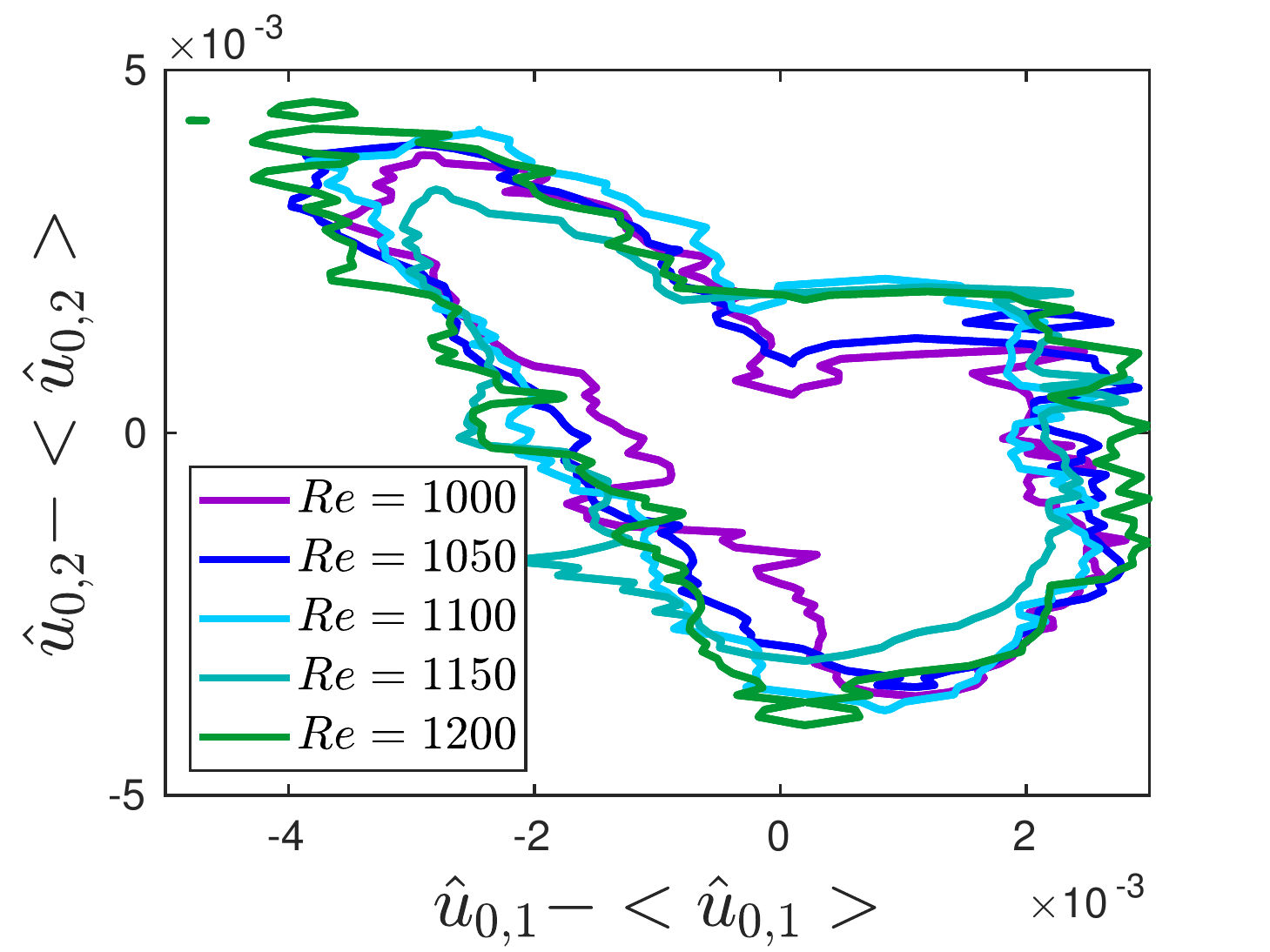}
    \label{fig:instanton_pdf_Re_split}}    
    \caption{Contours of the joint PDF of $(E_x,E_z)$ for successful trajectories going from (a) $\setAone$ to $\setBo$ or from (b) $\setAone$ to $\setBtwo$ at different Reynolds numbers. The temporal average during the transient trajectories is subtracted for better comparison, since sets $\setAone$ and $\setBtwo$ evolve with $Re$. (a) and (b) show the contours of the joint PDFs of $(E_x, E_z)$ and of $(\hat{u}_{0,1}, \hat{u}_{0,2})$ enclosing $90\%$ of the total probability.}
    \label{fig:instanton_pdf_Re}
\vspace*{-0.45cm}
\end{figure}

Figure 11 of the main article illustrates reactive tubes corresponding to decay (at $Re=830$) and to splitting (at $Re=1150$). The reactive tubes are isosurfaces of the probability density obtained from reactive trajectories going from $\setA$ to $\setBo$ or $\setBtwo$.
Here we investigate the effect of $Re$ on these reactive tubes. Figures \ref{fig:instanton_pdf_Re_decay} and \ref{fig:instanton_pdf_Re_split} compare trajectory concentration at different $Re$ by showing the contours of the probability density obtained from reactive trajectories in the phase spaces $(E_x, E_z)$ (for decays) and $(\hat{u}_{0,1}, \hat{u}_{0,2})$ (for splits). The contours surround $90\%$ of the probability. These plots are 2D projections of Figure 11.  

For decay cases,  the reactive tubes seem to contract slightly during the final viscous phase of the decay process as $Re$ is increased and decay becomes rarer.
In the case of splits, portions of the reactive tubes contract as $Re$ is decreased.
These plots indicate that the reactive trajectories become slightly more concentrated as $Re$ approaches $Re_c$. However, the range of $Re$ under study is restricted. It would be helpful to have data for decay events at $Re > Re_c$ and splits for $Re < Re_c$, both of which are still out of reach in our computations. 

\section{Approach to an edge state during decays}

\begin{figure}[t]
    \centering
    \subfloat[]{\includegraphics[width=0.5\textwidth]{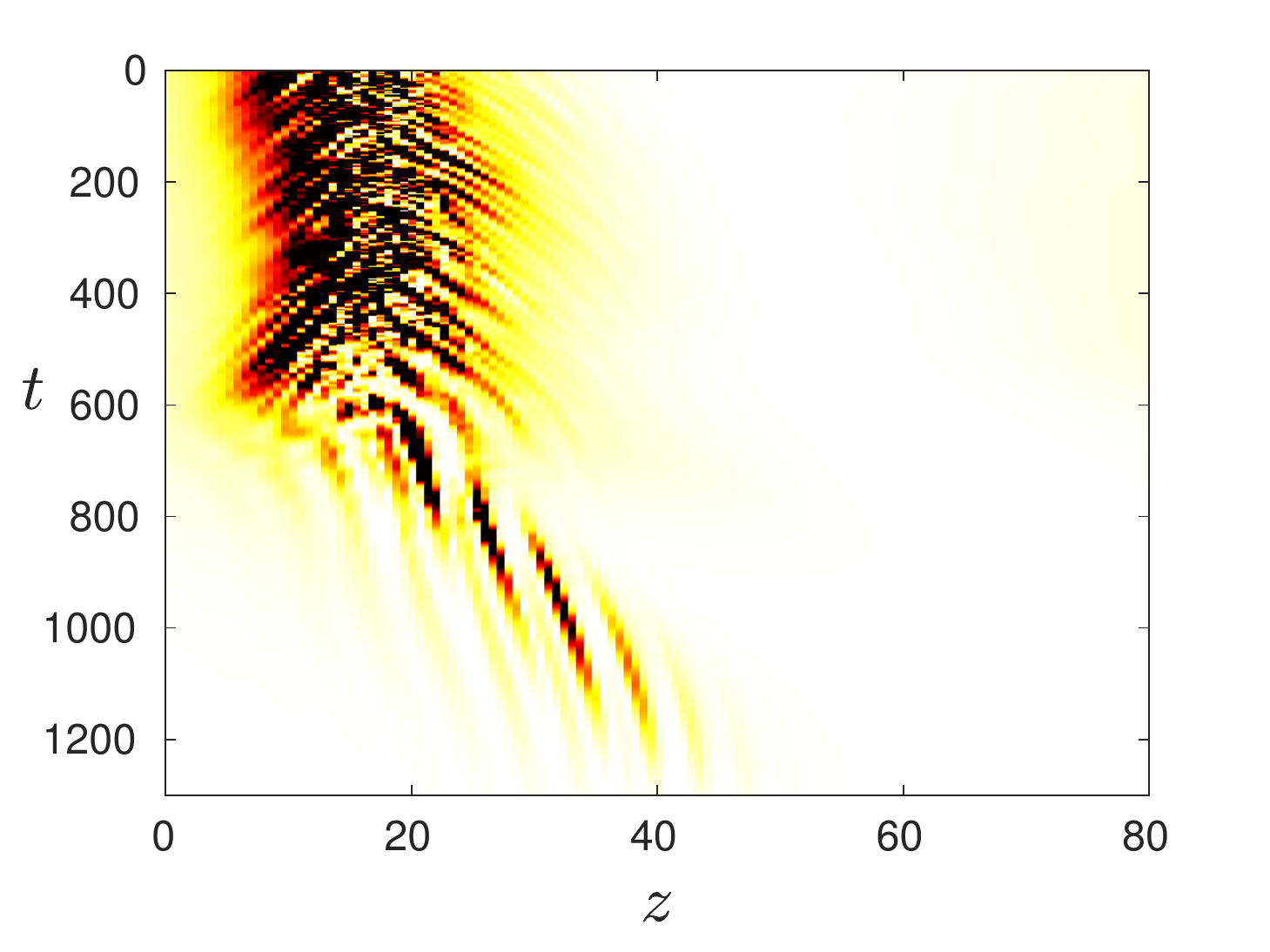} \label{fig:edge_probes}} 
    \subfloat[]{\includegraphics[width=0.5\textwidth]{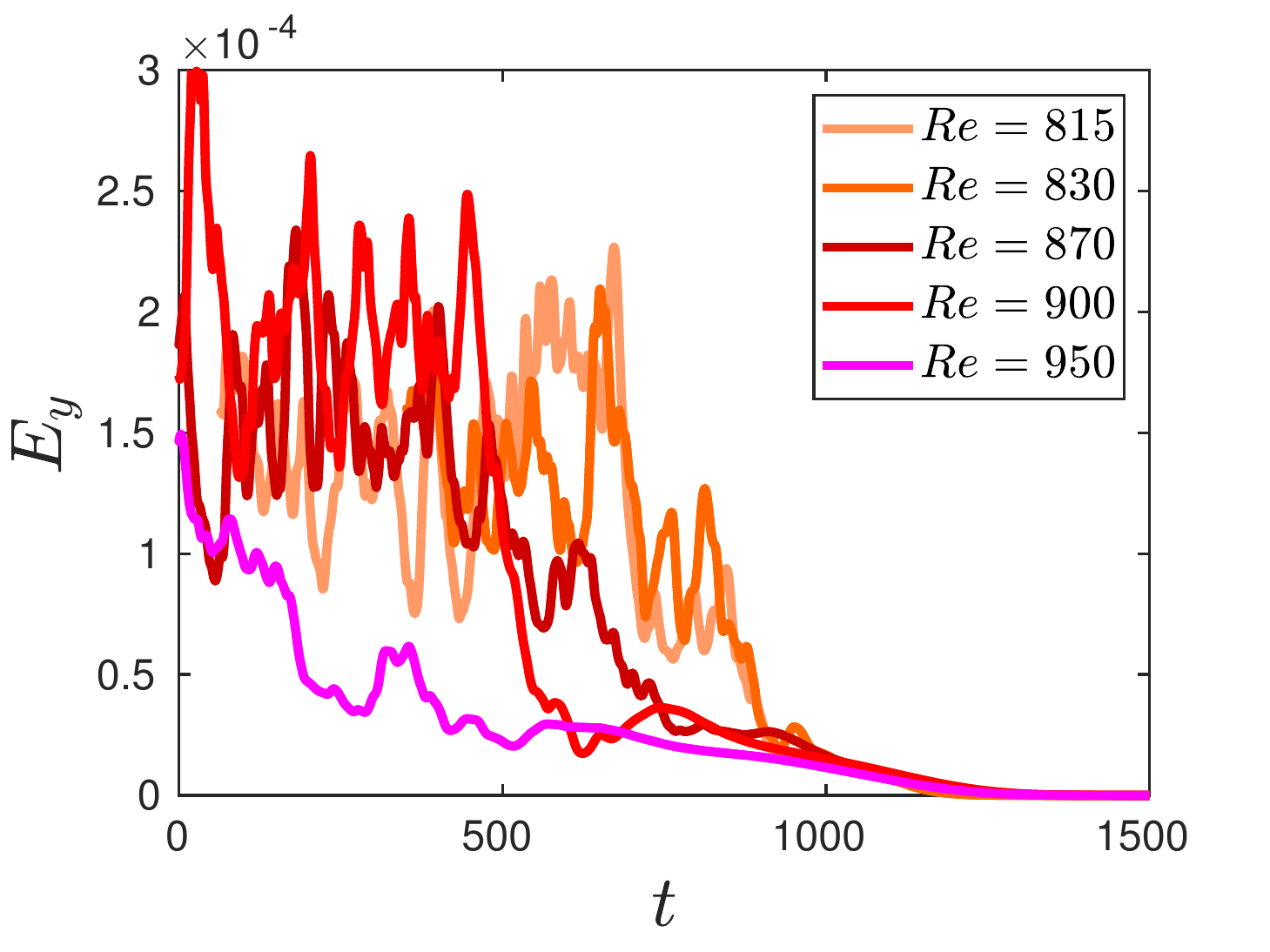}
    \label{fig:edge_v2_w2}} \\[-0.08cm]
    \subfloat[]{\includegraphics[width=\textwidth]{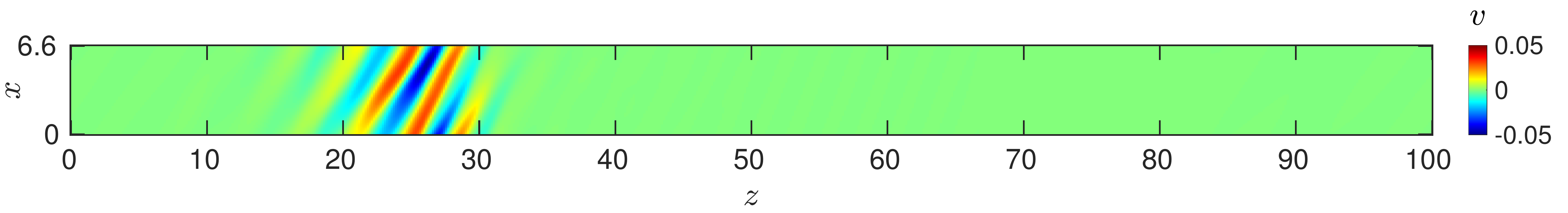}
    \label{fig:edge_state_xz}}\\
\vspace*{-13pt}    
     \subfloat[]{\includegraphics[width=\textwidth]{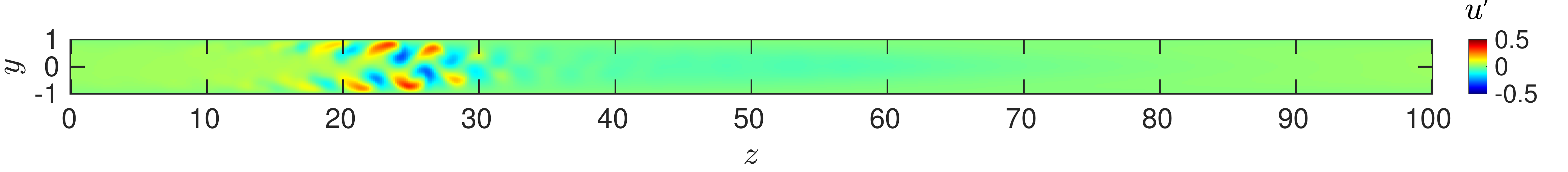}
    \label{fig:edge_state_yz}}\\[-0.1cm]
    \caption{Approach to the edge state of Paranjape \emph{et al.}  \cite{paranjape2020oblique}: (a) Space-time diagram of the decay of a turbulent band at $Re = 900$. Colors (white: 0, black: 0.001) show the local deviation energy at $(x=3.3,y=0.8)$. (b) 
    $E_y(t)$ for decay paths for different $Re$. 
    Stagnation indicating proximity to an edge state is particularly noticeable at $Re=900$ and $950$. Trajectory at $Re=900$ corresponds to (a).
    (c, d) Visualisation of possible edge state in the $(z,x)$ plane (cross-flow velocity $v$) and the $(z,y)$ plane (streamwise velocity $u'$), corresponding to $t=800$ in the space-time plot.  
    }
\end{figure}

The question of whether a saddle-point effectively separates the phase space between $\setAone$ and $\setBo$ or $\setBtwo$ can be answered by bisection techniques \cite{schneider2007turbulence, duguet2008transition}, as was done by Paranjape \emph{et al.} \cite{paranjape2020oblique} between one band and the laminar state. The computation of multiple successful trajectories also helps to verify the presence of this edge state, that should be statistically approached by reactive trajectories.
We show in Figure \ref{fig:edge_probes} a typical spatio-temporal diagram in the parameter range $Re \in [900, 950]$: during the decay of the band and before its full laminarisation, the trajectory approaches a state composed of weak straight streaks that differs from the one-band state. This state is visualised in Fig.~\ref{fig:edge_state_xz} and \ref{fig:edge_state_yz} and resembles the edge state found by Paranjape \emph{et al.}.
As shown by Fig.~\ref{fig:edge_probes}, this state moves at a velocity that differs from that of the initial turbulent band, and is approached within a time window of around 600 time units starting from $t=600$. 
The presence of an edge state is supported by Figure \ref{fig:edge_v2_w2}, which shows $E_y(t)$ for decaying trajectories. 
Proximity to the edge state is seen for $Re=900$ and $Re=950$ as approximate stagnation before the viscous decay. The particular case of $Re=900$ (red curve in Fig.~\ref{fig:edge_v2_w2}, and space-time diagram in Fig.~\ref{fig:edge_probes}) exemplifies a characteristic three-step process:  a first departure from the initial one-band state ($t\simeq 450$), followed by an approach to a plateau ($t\simeq 600$) correlated to the appearance of straight streaks (Fig.~\ref{fig:edge_probes}), which eventually decay  exponentially ($t\gtrsim1000)$. For $Re \leq 830$, the energy decays directly from the one-band state to the laminar state and the plateau does not appear. The stagnation phase, which differs from the subsequent exponential decay, confirms the nonlinear nature of the dynamics in this region, and suggests that we are near the edge state computed by Paranjape \emph{et al.} \cite{paranjape2020oblique}.

Our simulations support the established idea that pathways are statistically mediated by an underlying edge state when transiting from the one-band state to laminar flow, and that the system remains longer near the saddle point when the transition probability is lower (or $Re$ increased: see the longer stagnation phase at $Re=950$ than at $Re=900$). The importance of the edge state at higher $Re$ is consistent with the higher concentration observed on Fig.~\ref{fig:instanton_pdf_Re_decay} and with the longer reactive times (Fig.~12b).

\bibliographystyle{ieeetr}
\bibliography{bib}

\end{document}